\documentclass[a4paper,11pt]{article}
\usepackage{graphicx}
\usepackage{amsmath}
\usepackage{epsf}
\usepackage{epsfig}
\usepackage{amssymb}
\usepackage{epstopdf}
\usepackage{lmodern}
\usepackage{latexsym}
\usepackage{subfigure}
\usepackage{subfigure}
\usepackage{multirow}
\usepackage{txfonts}

\begin{document}
\begin{titlepage}
\vspace{.1cm}
\begin{center}

\vspace{.5cm}
{\bf\Large Flavour Physics in the Soft Wall Model}\\[.3cm]
\vspace{1cm}
Paul R.~Archer$^{a,}$\footnote{p.archer@sussex.ac.uk}, Stephan J.~Huber$^{a,}$\footnote{s.huber@sussex.ac.uk}
and 
Sebastian J\"{a}ger$^{a,}$\footnote{s.jaeger@sussex.ac.uk} \\ 
\vspace{1cm} {\em  
$^a$Department of Physics \& Astronomy, University of Sussex, Brighton
BN1 9QH, UK }\\[.2cm]

\end{center}
\bigskip\noindent
\vspace{1.cm}

\begin{abstract}
We extend the description of flavour that exists in the Randall-Sundrum (RS) model to the soft wall (SW) model in which the IR brane is removed and the Higgs is free to propagate in the bulk. It is demonstrated that, like the RS model, one can generate the hierarchy of fermion masses by localising the fermions at different locations throughout the space. However, there are two significant differences. Firstly the possible fermion masses scale down, from the electroweak scale, less steeply than in the RS model and secondly there now exists a minimum fermion mass for fermions sitting towards the UV brane. With a quadratic Higgs VEV, this minimum mass is about fifteen orders of magnitude lower than the electroweak scale. We derive the gauge propagator and despite the KK masses scaling as $m_n^2\sim n$, it is demonstrated that the coefficients of four fermion operators are not divergent at tree level. FCNC's amongst kaons and leptons are considered and compared to calculations in the RS model, with a brane localised Higgs and equivalent levels of tuning. It is found that since the gauge fermion couplings are slightly more universal and the SM fermions typically sit slightly further towards the UV brane, the contributions to observables such as $\epsilon_K$ and $\Delta m_K$, from the exchange of KK gauge fields, are significantly reduced.
\end{abstract}
\end{titlepage}
%%%%%%%%%%%%%%%%%%%%%%%%%%%%%%%%%%%%%%%%%%%%%%%%%%%%
\section{Introduction}

The soft wall (SW) model was originally proposed as a potential dual
to a field theory with linear confinement \cite{Karch:2006pv}, the idea being that under the AdS/CFT correspondence bound states in the, now broken, conformal field theory are conjectured to be dual to Kaluza-Klein (KK) modes on the gravity side \cite{Maldacena:1997re}. When, as in QCD, confinement is modelled such that the quark energy grows linearly with the quark separation then the mass of the $n$th meson will be subject to `Regge scaling', $m_n^2\sim n$. However, on the gravity side, when a discrete KK spectrum is obtained by imposing two hard cut offs on the space, for example by cutting off the space with two branes, then the KK masses will generically scale as $m_n^2\sim n^2$ \cite{Schreiber:2004ie, Shifman:2005zn}. 

In the SW model, the hard cut off in the IR of the theory is replaced
by a smooth cut off. To be specific, one considers an AdS${}_5$ space which is cut off in the UV by a brane but extends to infinity in the IR in an analogous fashion to the the RSII model \cite{Randall:1999vf}. However, in the IR the space has a cut off due to a smoothly decaying background value of a `dilaton' given by
\begin{equation}
\label{ }
S=\int d^5x \sqrt{g}e^{-\Phi}\mathcal{L}\quad\rm{where}\quad\Phi=\left (\frac{z}{R^{\prime}}\right )^\nu.
\end{equation}
Although we refer to this field as a dilaton, it should be stressed
this is very much a `bottom up' model. When $\nu>1$ this smooth cut
off gives rise to a discrete (bosonic) KK spectrum with \cite{Falkowski:2008fz, Batell:2008me}
\begin{displaymath}
m_n^2\sim \frac{n^{2-\frac{2}{\nu}}}{R^{\prime\; 2}}.
\end{displaymath} 
Hence with $\nu=2$ one obtains a Regge scaling in the KK spectrum. For
the majority of this paper we shall focus on this  scenario, although it is worth pointing out that when $\nu<1$ one obtains a continuous KK spectrum, or in other words the propagator has no poles. Such a scenario is conjectured to be dual to unparticles. When $\nu=1$ one obtains a continuous KK spectrum above a mass gap.

In recent years there has been significant interest in the
phenomenological implications of AdS${}_5$ in the context of the
Randall and Sundrum (RS) model \cite{Randall:1999ee}. This is
principally because the model appears to offer a geometrical
explanation of the two apparent hierarchies of the standard model
(SM). In particular by localising the Higgs on the IR brane, the
electroweak scale is gravitionally redshifted down from the Planck
scale and hence the model offers a potential resolution to the gauge
hierarchy problem. Likewise, by allowing the SM fermions and gauge
fields to propagate in the bulk, one can use a bulk mass term to
localise the fermions towards the UV and away from the Higgs. Such a
scenario not only offers an explanation of the hierarchy in the
fermion masses \cite{Grossman:1999ra, Gherghetta:2000qt, Huber:2000ie}
but it also offers a suppression of potential flavour changing neutral currents (FCNC's) \cite{Huber:2003tu, Agashe:2004cp}. While this provides a compelling description of flavour it is found that a problem arises when one considers the heavier quarks. With the Higgs localised on the IR brane the possible fermion zero mode masses scale down from the electroweak scale and hence the heavier quarks typically have to sit quite far towards the IR brane and in a region where FCNC's are no longer suppressed. This results in constraints, in particular from $K^0-\bar{K}^0$ mixing, forcing the mass of the first KK gauge mode to be $\gtrsim 20-30$ TeV \cite{Csaki:2008zd, Blanke:2008zb, Bauer:2009cf}.

From the holographic perspective the RS model is conjectured to be
dual to a theory which is closely related to technicolor \cite{
  ArkaniHamed:2000ds, Rattazzi:2000hs, PerezVictoria:2001pa}, albeit
technicolor without linear confinement. So a natural question one may
ask is which of the phenomenological features of the RS model be
translated across to the SW model. For example, can the space be
stabilised such that the gauge hierarchy problem be resolved? When one
stabilises such spaces using the Goldberger-Wise mechanism
\cite{Goldberger:1999uk} one finds an IR curvature singularity
\cite{Gubser:2000nd, Aybat:2010sn, George:2011gs}. Hence with no IR
cut off, the SW model suffers from a naked singularity. In any valid
solution the boundary terms in the equations of motions must vanish at
this singularity \cite{Cabrer:2009we}. This provides an additional
constraint, on the SW model, which forbids the space from being cut
off too sharply. In particular it has been argued that a warp factor,
sufficient to resolve the gauge hierarchy problem, can be generated
when $\nu < 1$ \cite{Gherghetta:2010he} or $\nu < 2$
\cite{Cabrer:2009we}.
None the less it is interesting to note that Regge scaling, in the KK masses, occurs in a class of SW solutions at the limit of what can still potentially offer a resolution of the hierarchy problem. 

Of course generating a large hierarchy is only one aspect of resolving the gauge hierarchy problem. One must also demonstrate that electroweak symmetry can be broken such that corrections to electroweak observables do not force the KK scale to a level at which one reintroduces a so called `little hierarchy problem'. In the SW model, with the absence of an IR brane, the Higgs must necessarily propagate in the bulk. This gives rise to a $z$ dependent Higgs VEV, $h(z)\sim h_0z^\alpha$. If such a Higgs VEV is too flat then it is found that one requires fine tuning in order to generate the correct electroweak scale, in particular it is found that to avoid this tuning $\alpha \gtrsim 2$ \cite{Cabrer:2011fb}. However a bulk Higgs also leads to a suppression in the couplings between KK gauge fields and the Higgs, which in turn gives rise to a significant reduction in the constraints coming from electroweak observables \cite{Falkowski:2008fz, Cabrer:2011fb, Cabrer:2010si, Carmona:2011ib}. The extent to which electroweak constraints are suppressed is sensitive to how flat the Higgs VEV is although, with a nearly quadratic Higgs VEV, it is possible to reduce the constraints to $M_{\rm{KK}}\gtrsim 1$ TeV without a custodial symmetry \cite{Cabrer:2011fb}.   

The focus of this paper is to examine whether or not the RS
description of flavour can be translated to models with a SW: Can the
model generate a hierarchy of fermion masses without a hierarchy in
the Yukawa couplings, and to what extent are FCNC's suppressed? Before
one can carry out a meaningful study there are a number of issues that
must be addressed. Firstly, as explained in \cite{Delgado:2009xb,
  Gherghetta:2009qs, MertAybat:2009mk}, when the fermions are
propagating in the bulk, the smooth cut off provided by the dilaton is
not sufficient to give rise to a discrete KK spectrum. In order to
ensure the latter, one must  
couple the fermions to something that is gaining a $z$ dependent
VEV. A natural candidate for this (used in \cite{Delgado:2009xb,
  Gherghetta:2009qs, Atkins:2010cc}) is the Higgs. However here,
following \cite{MertAybat:2009mk}, we introduce a generic $z$
dependent mass term that could arise from the Yukawa couplings to the
Higgs or some other new physics (such as couplings to the dilaton or
a Goldberger-Wise scalar \cite{Medina:2010mu}). It should be stressed
that such a $z$-dependent term must be present in any realistic soft wall
model with bulk fermions. We then proceed to demonstrate that a hierarchy of fermion masses can be generated when $\alpha\gtrsim 2$.
  
FCNC's occur in such a model due to the tree level exchange of KK
gauge fields. The standard approach to computing the size of such
FCNC's is to integrate out the KK modes in order to obtain the
coefficients of four fermion operators. The computation of such
coefficients will include a sum over the KK modes. Naively one may be
concerned that, with $m_n^2\sim n$, such a sum would be
logarithmically divergent, i.e. such processes would be dominated by
the higher KK modes. It is generally thought, however, that models
such as the present one are effective theories and unknown UV
dynamics will cut off any such divergence. Either way,
in section \ref{sect:Prop}, the gauge propagator is computed and it is
demonstrated that, provided the fermion bulk mass is 
sufficiently $z$-dependent, this divergence does not occur at tree level.

In section \ref{sect:Lept} we compute the tree level corrections to
certain flavour violating lepton decays. This study allows us to check
the validity of our approximations against earlier work
\cite{Atkins:2010cc} and examine some of the central physics. Finally,
in section \ref{quarksector}, we apply our results to a study of the
quark sector. It should be stressed that this can not be considered a
complete study. Firstly we do not consider SW models in generality but
focus on a specific model with a quadratic Higgs VEV ($\alpha=2$) and
a quadratic dilaton (i.e. Regge scaling in the KK spectrum,
$\nu=2$). This model has been chosen since it is believed to be close
to the optimal resolution of the gauge hierarchy problem, i.e. it is
still able to generate the large hierarchy but with minimal
constraints coming from the electroweak precision tests \cite{Cabrer:2011fb}. It also allows for a reasonable level of analytical control over the calculation. Secondly we only consider the tree level contribution to gauge mediated FCNC's in the kaon sector. As mentioned, some of the tightest constraints on the RS model come from the contribution to $\epsilon_K$ coming from the exchange of KK gluons. Hence here we primarily focus on the kaon sector. It is found that, due to the Higgs propagating in the bulk, all the SM fermion zero modes can sit further towards the UV brane than in the RS model and hence one finds that the contribution to $\epsilon_K$ is much smaller in the SW model. So although we focus on the kaon sector we anticipate that the same effect would cause an increased suppression in all FCNC's compared to that of the RS model, with the Higgs localised on the IR brane. However, as found in \cite{Agashe:2008uz}, one would anticipate that an analogous suppression of FCNC's would be present in the RS model with a bulk Higgs. Although we also find a slight increase in the universality, of the couplings between KK gauge fields and UV localised fermions, which is caused by the removal of the IR brane. We conclude in section \ref{sect:conclusion}.                                                                                                                                                          

%%%%%%%%%%%%%%%%%%%%%%%%%%%%%%%%%%%%%%%%%%%%%%%%%%%%
\section{Obtaining the Fermion Mass Hierarchy in the Softwall Model} \label{sect:Ferm}

In extending the RS description of flavour to the SW model one faces a
number of additional complications related to the backreaction of the
Higgs on the fermion profiles. In the RS model, studies of flavour
typically involve large scans over many possible Yukawa couplings
\cite{Csaki:2008zd, Blanke:2008zb, Bauer:2009cf, Agashe:2006iy,
  Casagrande:2008hr}. In the SW model, the fermion profiles receive
contributions from the Yukawa couplings and the bulk mass terms. Hence
the same scans over parameter space would be computationally
challenging. As explained in \cite{Batell:2008me, Delgado:2009xb,
  MertAybat:2009mk}, an interesting feature of the SW model is that
one must couple the fermions to the Higgs VEV, or some $z$ dependent
mass quantity, in order to obtain a discrete KK spectrum and so one
cannot simply neglect such backreactions. In this paper, following
\cite{MertAybat:2009mk,Aybat:2009zd}, we shall employ a bulk mass
term consisting of a constant and a $z$-dependent part.
Fortunately we find that,
when the fermions are localised towards the UV, deformations in the
fermion profile brought about by changes to the constant part of the (effective)
bulk mass terms are far larger than those caused by changes in the
$z$-dependent part. Hence in this paper we shall assume the $z$
dependence as flavour-independent, giving a universal correction to
the fermion profiles.
The validity of this approximation is checked in this paper. In this section we shall clarify our
assumptions,
compute the fermion profiles and demonstrate when one can generate a fermion mass hierarchy.         

\subsection{The Fermion Profile}
Let us begin by considering the SM fermions propagating in AdS$_5$, 
\begin{equation}
\label{metric}
ds^2=\left (\frac{R}{z}\right )^2\left (\eta^{\mu\nu}dx_\mu dx_\nu-dz^2\right ),
\end{equation}
where $\eta^{\mu\nu}=\rm{diag}(+---)$. The space runs from a UV brane at $z=R$ to infinity but is dynamically cut off by the background value of the dilaton,
\begin{equation}
\label{ }
S=\int d^5x \sqrt{g}e^{-\Phi}\mathcal{L}\quad\rm{where}\quad\Phi=\frac{z^2}{R^{\prime\;2}}.
\end{equation}
As mentioned in the introduction here we shall focus on a quadratic dilaton that yields Regge scaling in the KK masses. In this model the KK scale will be given by $M_{\rm{KK}}=\frac{1}{R^{\prime}}$, while the 4D effective Planck mass will be of the order $M_{\rm{Pl}}\sim\frac{1}{R}$. Hence, in order to offer a potential resolution to the gauge hierarchy problem  one requires a warp factor of
\begin{displaymath}
\Omega\equiv\frac{R^{\prime}}{R}\sim 10^{15}.
\end{displaymath} 
Here we shall not consider the stability of the space but simply impose this warp factor by hand and refer readers to \cite{Cabrer:2009we} for specific realisations of this model. In order to yield a low energy chiral SM it is necessary to include both doublets ($\Psi$) and singlets ($\Upsilon$) under $SU(2)$. If one couples the fermions, with a Yukawa coupling $Y$, to a bulk Higgs (assumed to be a $SU(2)$ doublet) that gains a VEV $h(z)$ then a single generation is described by
\begin{eqnarray}
S=\int d^5x\; \sqrt{g}e^{-\Phi}\Bigg [ \frac{1}{2}(i\bar{\Psi}\Gamma^M\nabla_M\Psi-i\nabla_M\bar{\Psi}\Gamma^M\Psi)-M_\Psi\bar{\Psi}\Psi\hspace{2cm}\nonumber\\
+\frac{1}{2}(i\bar{\Upsilon}\Gamma^M\nabla_M\Upsilon-i\nabla_M\bar{\Upsilon}\Gamma^M\Upsilon)
-M_\Upsilon\bar{\Upsilon}\Upsilon+Y( h\bar{\Psi}\Upsilon+\rm{h.c.})\Bigg ],
\end{eqnarray}
where $\Gamma^M=E^M_A\gamma^A$, $E_A^M$ is the F\"{u}nfbein, $\gamma^A$ are the Dirac matrices, $\nabla_M=D_M+\omega_M$ is the covariant derivative and $\omega_M$ is the spin connection. The upper latin indices run over all five dimensions, while $\mu,\;\nu$ run from $0$ to $3$. Since in five dimensions fermions are not chiral, one can include a bulk mass term $M$. 

We decompose the fermion field into two Weyl fermions such that $\Psi=\psi_L+\psi_R$ and $\gamma^5\psi_{L,R}=\mp\psi_{L,R}$ and then make the KK decompositions
\begin{equation}
\label{ FermKKdeomp}
\Psi_{L,R}=\sum_n \frac{z^2}{R^2}e^{\Phi/2}f^{(n)}_{\Psi\;L,R}(z)\Psi_{L,R}^{(n)}\quad\mbox{   and   }\quad\Upsilon_{L,R}=\sum_n \frac{z^2}{R^2}e^{\Phi/2}f^{(n)}_{\Upsilon\;L,R}(z)\Upsilon_{L,R}^{(n)},
\end{equation}      
requiring
\begin{equation}
\label{SWorthogOfDiag}
\int^{\infty}_R dz \left (f_{\Psi\;L,R}^{(n)}f_{\Psi\;L,R}^{(m)}+f_{\Upsilon\;L,R}^{(n)}f_{\Upsilon\;L,R}^{(m)}\right )=\delta^{nm}.
\end{equation}
If one defines the four dimensional effective mass by the Dirac equation $i\gamma^\mu\partial_\mu\psi^{(n)}_{L,R}=m_n\psi^{(n)}_{R,L}$, then the equations of motion are given by \cite{Batell:2008me, Gherghetta:2009qs, Atkins:2010cc}
 \begin{equation}
\label{offDiagSWferms}
\pm\partial_z\left(\begin{array}{c}f_{\Psi\;R,L}^{(n)} \\f_{\Upsilon\;R,L}^{(n)}\end{array}\right)+\frac{R}{z}\left(\begin{array}{cc}M_{\Psi} & Y h(z) \\ Y h(z)& M_{\Upsilon}\end{array}\right)\left(\begin{array}{c}f_{\Psi\;R,L}^{(n)} \\f_{\Upsilon\;R,L}^{(n)}\end{array}\right)=m_n\left(\begin{array}{c}f_{\Psi\;L,R}^{(n)} \\f_{\Upsilon\;L,R}^{(n)}\end{array}\right).
\end{equation}
It is now possible to see the double role of the Higgs. Firstly it is
giving mass to the fermion zero modes which are associated with the SM
particles. Secondly, in order to arrive at a discrete fermion KK
spectrum, it is necessary for (\ref{SWorthogOfDiag}) to be convergent
and hence the fermion profiles must go to zero as
$z\rightarrow\infty$. Here this is achieved by coupling the fermions
to the Higgs such that $\lim_{z\rightarrow\infty}\frac{h(z)}{z} = \infty$ \cite{Gherghetta:2009qs}. One can also see the difficulties that would arise in scanning over many anarchic Yukawa couplings, since for each $3\times 3$ Yukawa matrix one would have to solve the six coupled differential equations and normalise all the solutions.

Hence here we shall follow \cite{MertAybat:2009mk} and introduce a $z$ dependent, flavour-diagonal, bulk mass term   
\begin{equation}
\label{ }
S=\int d^5x \sqrt{g}e^{-\Phi}\left (\bar{\Psi} (i\Gamma^M\nabla_M-M(z)) \Psi \right ),
\end{equation} 
where
\begin{equation}
\label{bulkmass}
M(z)=\frac{c_0}{R}+\frac{c_1}{R}\frac{z^\alpha}{R^{\prime\;\alpha}}.
\end{equation}
There are two possible interpretations of such a mass term. One
possibility is to view such a mass term as arising from diagonalising
(\ref{offDiagSWferms}), and hence one would anticipate that there
would be nontrivial relationships between $c_0$ and $M_{\Psi
  /\Upsilon}$ and between $c_1 z^\alpha$ and $Yh(z)$. This equivalence
has been discussed in \cite{Gherghetta:2009qs}.
Alternatively one could attribute the term to some new high scale
physics, in which case the exponent of the mass term $\alpha$ need not
be the same as that of the Higgs VEV. Here we shall try to remain as
open minded as possible as to the underlying physics, keeping in mind
that,
in order to arrive at a discrete fermion spectrum, $c_1$ cannot be zero and hence any realistic scenario must give rise to such a term. Through a slight abuse of notation, we shall refer to the $c_0$ term as a bulk mass parameter and the $c_1$ term as the term arising from the Yukawa coupling. The advantage with introducing such a mass term is one can now find analytical expressions for the fermion profiles. In particular after making the same KK decomposition (\ref{ FermKKdeomp}) such that\begin{equation}
\label{orthogferm}
\int_R^\infty dz\, f_{L,R}^{(n)}f_{L,R}^{(m)}=\delta_{nm},
\end{equation} 
and the equations of motion are now
\begin{equation}
\label{ fermionsODE}
\left (\partial_z \pm \frac{R}{z}M(z)\right )f_{L,R}^{(n)}=\pm m_nf_{R,L}^{(n)},
\end{equation}
where the $\pm$ act on $f_L$ and $f_R$ respectively. The zero mode profile ($m_n=0$) can now be solved for
\begin{equation}
\label{fermionzeromode}
f_{L,R}^{(0)}=\sqrt{\frac{\frac{\pm 2c_1}{(R^{\prime})^{1\mp 2c_0}}\left (\frac{\alpha}{\pm 2c_1}\right )^{1-\frac{1\mp 2c_0}{\alpha}}}{\Gamma\left (\frac{1\mp 2c_0}{\alpha},\pm \frac{2c_1}{\alpha}\Omega^{-\alpha}\right )}}\;z^{\mp c_0}\exp \left (\mp\frac{c_1}{\alpha}\frac{z^\alpha}{R^{\prime\;\alpha}}\right ),
\end{equation} 
where $\Gamma(a,x)=\int_x^\infty e^{-t}t^{a-1}dt$ is the incomplete gamma function. Note the zero mode only exists if $\frac{c_1}{\alpha}>0$ for $\psi_L$ (and $\psi_R$ has UV Dirichlet boundary conditions) or $\frac{c_1}{\alpha}<0$ for $\psi_R$ (and $\psi_L$ has UV Dirichlet boundary conditions). Also, as in the RS case, the $\psi_L(\psi_R)$ profile will sit towards the UV when $c_0>\frac{1}{2}$ $(c_0<-\frac{1}{2})$. To find the profiles of the KK fermions one can combine (\ref{ fermionsODE}) to obtain
\begin{displaymath}
\left (\partial_z^2\mp \left (\frac{c_0}{z^2}+\frac{(1-\alpha)c_1}{z^2}\frac{z^\alpha}{R^{\prime\;\alpha}}\right )-\frac{c_0^2}{z^2}-\frac{2c_0c_1}{z^2}\frac{z^\alpha}{R^{\prime\;\alpha}}-\frac{c_1^2}{z^2}\frac{z^\alpha}{R^{\prime\;\alpha}}+m_n^2\right )f_{L,R}=0.
\end{displaymath}  
Imposing boundary conditions, obtained from requiring (\ref{
  fermionsODE}) to allow a zero mode, this can be analytically solved when $\alpha =2$ to give \cite{MertAybat:2009mk}
\begin{eqnarray}
f_L^{(n)}= \begin{cases}
 N\exp \left (\frac{-c_1z^2}{2R^{\prime\; 2}}\right )z^{-c_0}U\left (-\frac{R^{\prime\;2}m_n^2}{4c_1},\frac{1}{2}-c_0,\frac{c_1z^2}{R\prime\;2}\right )\quad\quad\hspace{1.23cm}\mbox{ for } c_1>0 \\
-N\frac{m_n}{2}\exp \left (\frac{c_1z^2}{2R^{\prime\; 2}}\right )z^{1+c_0}U\left (1+\frac{R^{\prime\;2}m_n^2}{4c_1},\frac{3}{2}+c_0,\frac{-c_1z^2}{R\prime\;2}\right )\quad\quad\mbox{ for } c_1< 0
 \end{cases}
 \nonumber\\
f_R^{(n)}= \begin{cases}
 N\frac{m_n}{2}\exp \left (\frac{-c_1z^2}{2R^{\prime\; 2}}\right )z^{1-c_0}U\left (1-\frac{R^{\prime\;2}m_n^2}{4c_1},\frac{3}{2}-c_0,\frac{c_1z^2}{R\prime\;2}\right )\quad\quad\hspace{0.3cm}\mbox{ for } c_1>0 \hspace{0.08cm}\\
 N\exp \left (\frac{c_1z^2}{2R^{\prime\; 2}}\right )z^{c_0}U\left (\frac{R^{\prime\;2}m_n^2}{4c_1},\frac{1}{2}+c_0,\frac{-c_1z^2}{R\prime\;2}\right )\quad\quad\hspace{1.55cm}\mbox{ for }  c_1< 0
 \end{cases}\label{KKfermionModes}
\end{eqnarray}
where $U(\alpha, \beta, x)$ are confluent hypergeometric functions or Kummer functions. The KK masses of these fields have been included in table \ref{Tab:FermMasses} and are in agreement with \cite{MertAybat:2009mk}.

\subsection{The Fermion Masses}

Having obtained the fermion zero mode profile we will treat the fermion mass, arising from the coupling to the Higgs, as a perturbation. I.e. we let the fermion zero mode masses be approximated by 
\begin{equation}
\label{MassIntegral}
M_{ij}=\int_R^{\infty}dz\frac{R}{z}Y_{ij}h(z)f_{L}^i(z)f_R^j(z),
\end{equation}     
where $i,j$ are flavour indices and $f_{L,R}^i$ is the zero mode profile (\ref{fermionzeromode}) with $c_0=c_0^{L,R\;i}$ and $c_1=c_1^{L,R\;i}$. Here we parameterise the Yukawa couplings by $Y_{ij}=\lambda_{ij}\sqrt{R}$, where $\lambda_{ij}$ are taken to be complex and order one. The Higgs VEV is assumed to be of the form \cite{Batell:2008me}
\begin{equation}
\label{higgsvev}
h(z)=h_0R^{-\frac{3}{2}}\left (\frac{z}{R^{\prime}}\right )^\alpha,
\end{equation}
where $h_0$ is a dimensionless constant.  If one assumes that the $c_1$ term in (\ref{bulkmass}) has arisen from the Higgs VEV then there will be a non trivial relationship between the $c_1^{L,R\;i}$ values and $h_0\lambda_{ij}$. Since the only way to know such a relationship would be to solve (\ref{offDiagSWferms}), one would wonder whether the situation has been improved. However fortunately it is found that, when the fermions are sitting towards the UV brane, the four dimensional effective couplings, that govern the low energy phenomenology, are far more sensitive to changes in the $c_0$ bulk mass parameters than changes in the $c_1$ parameters. This allows us to make the approximation of assuming a universal $c_1$ value, i.e. assuming that $c_1^{Li}=-c_1^{Ri}=c_1$. 

Turning back to the fermion masses (\ref{MassIntegral}) and substituting in (\ref{fermionzeromode}) gives
\begin{equation}
\label{ } 
M_{ij}=h_0\frac{\lambda_{ij}}{R^{\prime}}\frac{2^{(\frac{1-c_0^{Li}+c_0^{Rj}}{\alpha})}(c_1^{Li})^{\frac{1-2c_0^{Li}}{2\alpha}}(-c_1^{Rj})^{\frac{1+2c_0^{Rj}}{2\alpha}}}{\alpha^{\frac{1-\alpha}{\alpha}}(c_1^{Li}-c_1^{Rj})^{\frac{\alpha-c_0^{Li}+c_0^{Rj}}{\alpha}}}\frac{\Gamma\left (\frac{\alpha-c_0^{Li}+c_0^{Rj}}{\alpha},\frac{c_1^{Li}-c_1^{Rj}}{\alpha}\Omega^{-\alpha}\right )}{\sqrt{\Gamma\left (\frac{1-2c_0^{Li}}{\alpha},\frac{2c_1^{Li}}{\alpha}\Omega^{-\alpha}\right  )\Gamma\left (\frac{1+2c_0^{Rj}}{\alpha},\frac{-2c_1^{Rj}}{\alpha}\Omega^{-\alpha}\right  )}}.
\end{equation} 
Note that the integral in (\ref{MassIntegral}) is convergent only when $\alpha>0$, $(c_1^{Li}-c_1^{Rj})>0$ and $\frac{\alpha-c_0^{Li}+c_0^{Rj}}{\alpha}>0$. One can also see that the $c_0$ terms always appear in the exponent while the $c_1$ terms appear in the base. If one now assumes universal $c_1$ values then this expression simplifies to
\begin{equation}
\label{SWMasses}
M_{ij}=h_0\frac{\lambda_{ij}}{R^{\prime}}\left (\frac{\alpha}{2c_1}\right )^{\frac{\alpha-1}{\alpha}}\frac{\Gamma\left (\frac{\alpha-c_0^{Li}+c_0^Rj}{\alpha},\frac{2c_1}{\alpha}\Omega^{-\alpha}\right )}{\sqrt{\Gamma\left (\frac{1-c_0^{Li}}{\alpha},\frac{2c_1}{\alpha}\Omega^{-\alpha}\right )\Gamma\left (\frac{1+c_0^{Ri}}{\alpha},\frac{2c_1}{\alpha}\Omega^{-\alpha}\right )}}.
\end{equation}
This should be compared with the analogous expression for the RS model with the Higgs localised on the IR brane and $f_{(L,R)}^{(0)}=Ne^{\mp c_{L,R}kr}$ \cite{Huber:2000ie}\footnote{ Here we have used the metric $ds^2=e^{-2kr}\eta^{\mu\nu}dx_\mu dx_\nu-dr^2$ with $r\in [0,R]$ such that $\Omega\equiv e^{kR}$},
\begin{equation}
\label{RSMasses}
M_{ij}=\frac{\lambda_{ij}v\Omega}{k}f_L^i(R)f_R^j(R)=\lambda_{ij}v\sqrt{(1-2c_L^i)(1+2c_R^j)}\;\frac{\Omega^{1-c_L^i+c_R^j}}{\sqrt{(\Omega^{1-2c_L^i}-1)\;(\Omega^{1+2c_R^j}-1)}},
\end{equation}
where $v\approx 174$ GeV is the Higgs VEV.    

\begin{figure}[ht!]
    \begin{center}
        \subfigure[The RS Model]{%
            \label{fig:FermMassRS}
            \includegraphics[width=0.52\textwidth]{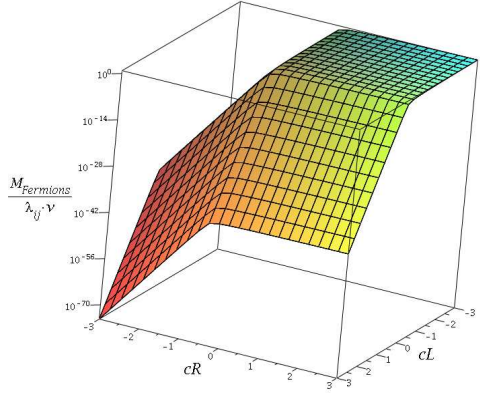}
        }
        \subfigure[$\alpha=1$]{%
           \label{fig:second}
           \hspace{-1.5cm}\includegraphics[width=0.52\textwidth]{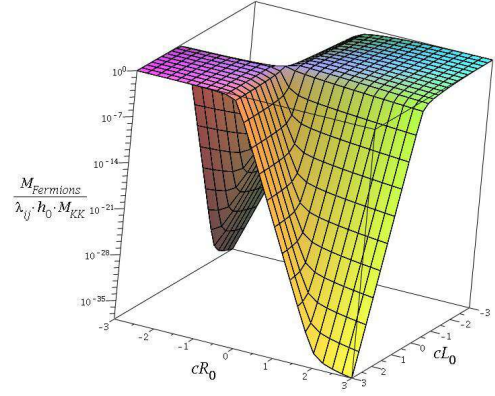}
        }\\ %  ------- End of the first row ----------------------%
        \subfigure[$\alpha=2$]{%
            \label{fig:FermMass2}
            \includegraphics[width=0.52\textwidth]{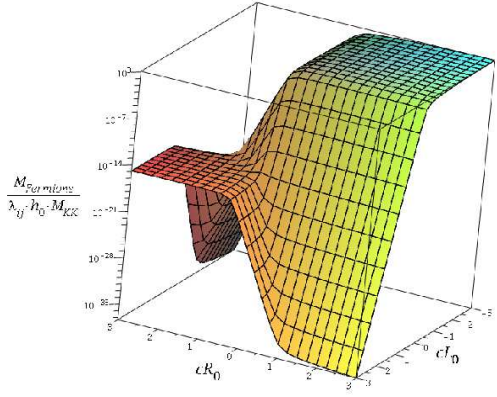}
        }%
        \subfigure[$\alpha=4$]{%
            \label{fig:FermMass4}
             \hspace{-1.5cm}\includegraphics[width=0.52\textwidth]{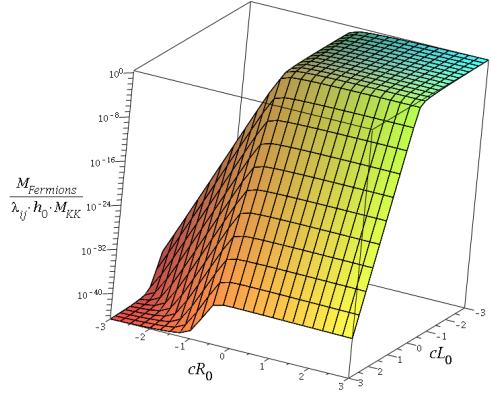}
        }%
   \label{fermmass}
    \end{center}
    \caption{The hierarchy of fermion masses for different Higgs VEV exponents. Here $c_1=1$ but it is found that the results are not sensitive to the value of $c_1$, while $c_{L,R}$ refer to the $c_0$ constant bulk mass terms. We take $\Omega=10^{15}$ and $M_{\rm{KK}}\equiv \frac{1}{R^{\prime}}=1\;\rm{TeV}$.  } \label{fermmass}
\end{figure}

The ranges of possible masses have been plotted in figure \ref{fermmass}. One can see two significant differences from the RS case. The first and most obvious, is that as one localises the fermions closer and closer to the UV one hits a minimum fermion mass. The location of this minimum is very sensitive to both the warp factor and the form of the Higgs VEV. In particular one would struggle to generate the hierarchy of fermion masses with a linear Higgs VEV or, for that a matter, a small warp factor. What is particularly interesting is that the optimum solutions to the gauge hierarchy problem, found in \cite{Cabrer:2011fb}, appear to correspond to $\Omega\approx 10^{15}$ and $\alpha\approx 2$. This would put the minimum fermion mass about fifteen orders of magnitude lower than the mass of an IR localised fermion. If one took $h_0M_{\rm{KK}}\sim 100$ GeV then this would correspond to a minimum fermion mass of the order $0.1$ meV which is roughly the scale of the observed neutrino masses. It is curious to note that this approximately corresponds to models which are found to have minimal electroweak constraints \cite{Cabrer:2011fb}. Hence the majority of this paper will focus on models with $\alpha=2$. One could obtain a lower mass by considering a split fermion scenario, in which the left-handed fermions are localised towards the opposite side of the space to the right-handed fermions. However one would suspect that this scenario would give rise to large FCNC's and hence would not be phenomenologically viable. Note that this potential relationship between the neutrino masses and the warp factor, in scenarios with a bulk Higgs, was first pointed out in \cite{Agashe:2008fe}.

The second difference, from the RS model, is the gradient of the
slope. In both the RS model and the SW model the possible masses drop
away exponentially. However, in the SW model, the exponent is no
longer constant. This leads us to anticipate small mass hierarchies for light fermions, such as neutrinos, and larger mass hierarchies between heavier fermions, such as the the top and bottom quarks. This reduced $c_0$ dependence also has implications in the suppression of FCNC's which we shall consider in later sections.

It should be noted that, although here we have treated the Yukawa couplings as a perturbation, these results are in good agreement with the corresponding plot in \cite{Gherghetta:2009qs} (with $\alpha=2$ and $\Omega=10^3$). One would also expect that these differences would be present in any model with a bulk Higgs (see for example \cite{Medina:2010mu}). Since this notion of a minimum in the 4D effective Yukawa coupling is generic to all couplings between fermion zero modes and profiles sitting towards the IR. It is, for example, well known that the same effect is present in the fermion couplings to KK gauge fields (see figure \ref{fig:FermCoupl}).

%%%%%%%%%%%%%%%%%%%%%%%%%%%%%%%%%%%%%%%%%%%%%%%%%%%%
\section{The Gauge Position / Momentum Space Propagator}\label{sect:Prop}
Having demonstrated that, with $\alpha\gtrsim 2$ and $\Omega=10^{15}$, one can generate the hierarchy of fermion masses we shall move on to look at FCNC's. In the RS model, the tree level exchange of KK gauge fields gives rise to FCNC's which provide some of the most stringent constraints on the model. The computations of such processes involve a sum over the KK tower ($\sim \sum_n\,1/m_n^2$) which, as already discussed, are potentially logarithmically divergent with $m_n^2\sim n$. Even if such processes are not divergent, due to the slower convergence with $n$, it is clearly necessary to include more KK modes than in the RS model in order to arrive at an accurate result, although it is not clear how many need to be included. With this in mind we shall now compute the 5D gauge propagator that inherently includes the full KK tower. For the remainder of this paper we shall restrict ourselves to studying models with a quadratic Higgs VEV ($\alpha=2$).

\subsection{The Photon Propagator}
We start by considering a $U(1)$ gauge field in the space (\ref{metric})
\begin{equation}
\label{guageaction}
S=\int d^5x \sqrt{g}e^{-\Phi}\left (-\frac{1}{4}F_{MN}F^{MN}\right ),
\end{equation}
where $M, N$ run over the five space time indices and $\mu, \nu$ run over the four large dimensions. Integrating by parts and introducing a gauge fixing term of the form\newline $\mathcal{L}_{GF}=-\frac{1}{2\xi}\frac{R}{z}e^{-\Phi} \left(\partial_\mu A^\mu -\xi e^{\Phi}\frac{z}{R}\partial_5 ( \frac{R}{z}e^{-\Phi} A^5) \right )^2$, one gets
\begin{eqnarray}
\label{ }
S=\int d^5x \bigg [ \frac{e^{-\Phi}}{2}\frac{R}{z}A_\mu\left (\eta^{\mu\nu}\partial^2-\left (1-\frac{1}{\xi}\right )\partial^\mu\partial^\nu-\frac{e^{\Phi}z}{R}\eta^{\mu\nu}\partial_5\left (e^{-\Phi}\frac{R}{z}\partial_5 \right )\right )A_\nu\nonumber\\
-\frac{e^{-\Phi}}{2}\frac{R}{z}A_5\partial^2A_5 +\frac{\xi e^{-\Phi}}{2}\frac{R}{z}A_5\;\partial_5\left (e^{\Phi}\frac{z}{R}\partial_5\left (e^{-\Phi}\frac{R}{z}A_5 \right )\right )\bigg ].
\end{eqnarray}
In this paper we will work at tree level in the unitary gauge ($\xi\rightarrow\infty$) and so neglect contributions from ghosts and the unphysical $A_5$ goldstone bosons. We refer the reader to \cite{Randall:2001gb} for a discussion of alternative gauges. After Fourier transforming with respect to the four large dimensions, such that $p_\mu=i\partial_\mu$, the gauge propagator is given by
\begin{equation}
\label{ }
\langle A^\mu A^\nu\rangle=-iG_p(z,z^{\prime})\left (\eta^{\mu\nu}-\frac{p^\mu p^\nu}{p^2}\right ),
\end{equation}
where
\begin{equation}
\label{Propag}
\left (e^{\Phi}\frac{z}{R}\partial_5\left (e^{-\Phi}\frac{R}{z}\partial_5 \right )+p^2 \right )G_p(z,z^{\prime})=e^{\Phi}\frac{z}{R}\delta(z-z^{\prime}).
\end{equation}
The point of all this is that in order to obtain a 4D effective theory one just has to integrate out the Green's function $G_p(z,z^{\prime})$. With a dilaton of the form $\Phi=\frac{z^2}{R^{\prime\; 2}}$, the most general solution of (\ref{Propag}) is
\begin{equation}
\label{ }
G_p(z,z^{\prime}) =
\begin{cases}
Az^2M(1-a,2,\frac{z^2}{R^{\prime\;2}})+Bz^2U(1-a,2,\frac{z^2}{R^{\prime\;2}}) & \text{if }z<z^{\prime} \\
 Cz^2M(1-a,2,\frac{z^2}{R^{\prime\;2}})+Dz^2U(1-a,2,\frac{z^2}{R^{\prime\;2}}) & \text{if }z>z^{\prime}
\end{cases}
\end{equation}
where $A,B,C,D$ are constants of integration. We have also introduced the quantity
\begin{displaymath}
a\equiv\frac{R^{\prime\; 2}p^2}{4},
\end{displaymath}
 while $M(\alpha,\beta,x)$ and $U(\alpha,\beta,x)$ are again Kummer functions\footnote{Here we use the notation of \cite{abramowitz+stegun}, although $M(\alpha,\beta,x)$ can be alternatively denoted by ${}_1F_1(\alpha,\beta,x)$ or $\Phi(\alpha,\beta,x)$, and likewise $U(\alpha,\beta,x)$ can be denoted $x^{-\alpha}{}_2F_0(\alpha,1+\alpha-\beta);\; ;-\frac{1}{x})$ or $\Psi(\alpha,\beta,x)$.}. Throughout the remainder of this paper we will repeatedly use relations taken from \cite{ abramowitz+stegun,Slater, Buchholz}. Following \cite{Randall:2001gb}, we introduce $u=\min(z,z^{\prime})$ and $v=\max(z,z^{\prime})$ and impose the continuity condition that matches the two solutions at $z=z^{\prime}$, i.e. $G_p(u,u)=G_p(v,v)$, obtaining
 \begin{eqnarray*}
 \normalsize
G_p(u,v)=Nu^2v^2\left (AM\left (1-a,2,\frac{u^2}{R^{\prime\;2}}\right )+BU\left (1-a,2,\frac{u^2}{R^{\prime\;2}}\right ) \right )\hspace{2cm}\\
\times\left (CM\left (1-a,2,\frac{v^2}{R^{\prime\;2}}\right )+DU\left (1-a,2,\frac{v^2}{R^{\prime\;2}}\right ) \right ).
\normalsize
\end{eqnarray*}
Here $N$ is a normalisation constant found by integrating over (\ref{Propag}) to get the `jump' condition
\begin{displaymath}
\lim_{\epsilon\rightarrow 0}\partial_zG_p(u,v)\lvert_{z^{\prime}-\epsilon}^{z^{\prime}+\epsilon}=\exp \left(\frac{z^{\prime\;2}}{R^{\prime\;2}}\right )\frac{z^{\prime}}{R}, 
\end{displaymath}
giving
\begin{displaymath}
N=\frac{\Gamma(1-a)}{2(BC-AD)RR^{\prime\;2}},
\end{displaymath}
where we have used the relation
\begin{displaymath}
\frac{e^x}{x\left (M(1-a,2,x)U(-a,2,x)+(1+a)U(1-a,2,x)M(-a,2,x)\right )}=\Gamma(1-a).
\end{displaymath}
To fix the constants of integration we impose Neumann boundary conditions on the UV brane, i.e. $\partial_uG_p(u,v)\vert_{u=R}=0$, to get
\begin{eqnarray}
A(a)=(\Omega^{-2}-a)U(1-a,2,\Omega^{-2})-U(-a,2,\Omega^{-2})\label{Aconst},\\
B(a)=(a-\Omega^{-2})M(1-a,2,\Omega^{-2})-(1+a)M(-a,2,\Omega^{-2})\label{Bconst}.
\end{eqnarray}
Since there is no IR brane, the determination of $C$ and $D$ is a little more subtle. Here we replace one of the boundary conditions with a `normalisability' condition that dictates that the propagator be comprised of KK modes which are normalisable with respect to $\int_R^{\infty}dz\;e^{-\Phi}\frac{R}{z}f_n^2=1$. This condition implies that
\begin{equation}
\label{IRNorm}
\int_R^{\infty}dz\;e^{-\Phi}\frac{R}{z}G_p(z,z)=\sum_n\int_R^{\infty}dz\;e^{-\Phi}\frac{R}{z}\frac{f_n^2(z)}{p^2-m_n^2}\sim\sum_n\frac{1}{n}.
\end{equation}
Hence we require that the integral (\ref{IRNorm}) be logarithmically divergent. For large $x$, Kummer functions scale as $M(\alpha,\beta,x)\sim\frac{\Gamma(\beta)}{\Gamma(\alpha)}e^{x}x^{\alpha-\beta}$ and $U(\alpha,\beta,x)\sim x^{-\alpha}$. The integrand will then scale as
\begin{displaymath}
\int_R^{\infty}dz\;e^{-\Phi}\frac{R}{z}G_p(z,z)\sim \int_R^{\infty}dz \left (AC e^{\frac{z^2}{R^{\prime\;2}}}z^{-4a-1}+(AD+BC)z^{-1}+BDe^{-\frac{z^2}{R^{\prime\;2}}}z^{4a-1}\right ).
\end{displaymath}
The last term is clearly convergent, and so with $A$ and $B$ already fixed, only by setting $C=0$ can one obtain a logarithmic divergence. This then results in $D$ being arbitrary and the full propagator being given by
\begin{equation}
\label{ }
G_p(u,v)=-\frac{\Gamma(1-a)u^2v^2}{2ARR^{\prime\;2}}\left (AM\left(1-a,2,\frac{u^2}{R^{\prime\;2}}\right )+BU\left (1-a,2,\frac{u^2}{R^{\prime\;2}}\right )\right )U\left (1-a,2,\frac{v^2}{R^{\prime\;2}}\right ).
\end{equation}
The KK masses will be given by the poles of the propagator, i.e. when $A(a)=0$. $G_p(u,v)$ will have the same form for the gluons, whereas the W/Z propagator will be slightly deformed.

\subsection{ The W/Z Gauge Fields}\label{sect:Zmass}
Before considering the W/Z propagator it is useful to first consider the individual KK modes. Working post spontaneous symmetry breaking, we add to (\ref{guageaction})  a mass term of the form
\begin{displaymath}
\Delta\mathcal{L}=\frac{1}{4}h(z)^2g^2A_\mu A^\mu,
\end{displaymath} 
where $g$ is the 5D coupling to the Higgs (or for the Z boson $\frac{1}{4}h(z)^2(g^2+g^{\prime\;2})A_\mu A^\mu$) and $h$ is the Higgs VEV (\ref{higgsvev}). Carrying out the usual KK decomposition, $A_\mu=\sum_n f_n(z)A_\mu^{(n)}(x^\mu)$, such that 
\begin{equation}
\label{GaugeOrthog}
\int_R^{\infty}dz\;e^{-\Phi}\frac{R}{z}f_nf_m=\delta_{nm},
\end{equation}
the gauge profile will then be given by
\begin{equation}
\label{Zeqn}
\left (\partial_5^2-\left(\frac{2z}{R^{\prime\;2}}+\frac{1}{z}\right )\partial_5-\frac{g^2h_0^2}{2Rz^2}\left(\frac{z}{R^{\prime}}\right )^{2\alpha}+m_n^2\right )f_n=0.
\end{equation}
When $\alpha=2$ this can be solved to give
\begin{equation}
\label{Zprofile}
f_n(z)=Nz^2\exp \left(\frac{1}{2}\frac{z^2}{R^{\prime\;2}}(1-\zeta)\right )\;U\left (1-\tilde{a}_n,2,\zeta\frac{z^2}{R^{\prime\;2}}\right ),
\end{equation}
where we have introduced the quantities
\begin{displaymath}
\zeta\equiv\sqrt{\frac{g^2h_0^2}{2R}+1} \quad\mbox{and}\quad\tilde{a}_n\equiv\frac{m_n^2R^{\prime\;2}}{4\zeta}. 
\end{displaymath}
Note on sending $g^2h_0^2\rightarrow 0$, $\zeta\rightarrow 1$ and bearing in mind that $U(\alpha,\beta,x)=x^{1-\beta}U(1+\alpha-\beta,2-\beta,x)$, one regains the gauge profiles for massless gauge fields found in \cite{Batell:2008me}. We have also imposed the normalisability condition (\ref{GaugeOrthog}) resulting in the coefficient in front of the $M(\alpha,\beta,x)$ part of the solution being set to zero. Imposing Neumann boundary conditions on the UV brane ($\partial_5f_n\lvert_{z=R}=0$) then gives 
\begin{displaymath}
\left ((1+\zeta)\Omega^{-2}-2\tilde{a}_n\right )U\left (1-\tilde{a}_n,2,\zeta\Omega^{-2}\right )-2U\left (-\tilde{a}_n,2,\zeta\Omega^{-2}\right )=0.
\end{displaymath}  
Once again as $\zeta\rightarrow 1$, the KK masses are the same as those given by the poles of the propagator (\ref{Aconst}). $\zeta$ can then be found by solving
\begin{displaymath}
\left ((1+\zeta)\Omega^{-2}-\frac{M_{W/Z}^2R^{\prime\;2}}{2\zeta}\right )U\left (1-\frac{M_{W/Z}^2R^{\prime\;2}}{4\zeta},2,\zeta\Omega^{-2}\right )=2U\left (-\frac{M_{W/Z}^2R^{\prime\;2}}{4\zeta},2,\zeta\Omega^{-2}\right ).
\end{displaymath}
With $M_{\rm{KK}}\approx 1-10$ TeV, then $\zeta_Z\approx 1.28-1.0028$ and hence when $M_{\rm{KK}}\ngg M_{W/Z}$ then $g^2\sim\mathcal{O}(R)$. In other words, provided the KK scale is not too large, one does not have to introduce any couplings that are very large or very small in order to generate the correct W and Z masses. 

We can now estimate $h_0$ by comparison with electroweak observables. Ideally one should compare to all observables in particular the Fermi constant. However, since one of the motivations for studying this model was its relatively small electroweak corrections, it is reasonable to just fit the gauge couplings and $h_0$ to three observables. In particular if one fits to the Z mass, fine structure constant, $\alpha$, and weak mixing angle, $\theta_w$, then
\begin{displaymath}
4\pi \alpha=g^2s_w^2f_0^2,
\end{displaymath}
where $s_w^2=\sin^2 \theta_w$ and $c_w^2=1-s_w^2$. We have also introduced the flat normalised photon gauge profile
\begin{equation}
\label{Fzero}
f_0=\sqrt{\frac{2}{R\;\mbox{E}_1(\Omega^{-2})}}\quad \mbox{and} \quad \mbox{E}_1(x)=\int_x^\infty dt\frac{e^{-t}}{t}.
\end{equation} 
Hence for a quadratic Higgs VEV we find
\begin{displaymath}
h_0^2\approx\frac{(\zeta_Z^2-1)c_w^2s_w^2}{\pi\alpha\; \mbox{E}_1(\Omega^{-2})}.
\end{displaymath}
By fitting the couplings of the five dimensional gauge field, $g$ and $g^{\prime}$, directly to observables we are assuming that the ratio of the 5D couplings is the same as the ratio of the 4D effective couplings. Hence we are neglecting $\mathcal{O}(M_{W/Z}^2/M_{\rm{KK}}^2)$ corrections to both the W and Z couplings and masses. One can check the validity of this approximation by numerically verifying that
\begin{displaymath}
\frac{\zeta_W^2-1}{\zeta_Z^2-1}=\hat{c}_w^2\approx c_w^2,
\end{displaymath}    
which gives $\hat{c}_w^2=0.7564,\;0.7716,\;0.7759$ and $0.7771$ for $M_{\rm{KK}}=1,\;2,\;4$ and $10$ TeV. Using these relations $h_0M_{\rm{KK}}\approx 262-245$ GeV for $\Omega=10^{15}$ and $M_{\rm{KK}}=1-10$ TeV. Having obtained $h_0$ one can now repeat the analysis of the previous section to obtain the W/Z propagator using
\begin{displaymath}
\left (\partial_5^2-\left(\frac{2z}{R^{\prime\;2}}+\frac{1}{z}\right )\partial_5-\frac{g^2}{Rz^2}\left(\frac{z}{R^{\prime}}\right )^{4}+p^2\right )G_p^{(W,Z)}(z,z^{\prime})=e^\Phi\frac{z}{R}\delta(z-z^{\prime}),
\end{displaymath}
to give
\begin{eqnarray}
G_p^{(W/Z)}(u,v)=-\frac{\zeta\Gamma(1-\zeta)u^2v^2\exp\left(\frac{1}{2}\frac{u^2}{R^{\prime\;2}}(1-\zeta)+\frac{1}{2}\frac{v^2}{R^{\prime\;2}}(1-\zeta)\right )}{2ARR^{\prime\;2}}\Bigg (AM\left (1-\tilde{a},2,\frac{\zeta u^2}{R^{\prime\;2}}\right )\nonumber\\
+BU\left (1-\tilde{a},2,\frac{\zeta u^2}{R^{\prime\;2}}\right )\Bigg)\;U\left (1-\tilde{a},2,\frac{\zeta v^2}{R^{\prime\;2}}\right ),
\end{eqnarray}
where $\tilde{a}\equiv\frac{p^2R^{\prime\;2}}{4\zeta}$ and 
\begin{eqnarray}
A(\tilde{a})=\left (\Omega^{-2}(1+\zeta)-2\tilde{a}\right )U\left (1-\tilde{a},2,\zeta \Omega^{-2}\right )-2 U\left (-\tilde{a},2,\zeta \Omega^{-2}\right ),\\
B(\tilde{a})=\left (2\tilde{a}-\Omega^{-2}(1+\zeta)\right )M\left (1-\tilde{a},2,\zeta\Omega^{-2}\right )-2(1+\tilde{a})M\left (-\tilde{a},2,\zeta\Omega^{-2}\right ).
\end{eqnarray}
As already mentioned, for the majority of this paper we shall focus on the quadratic Higgs VEV ($\alpha=2$). None the less it is worth pointing out the special case of a linear VEV. With $\alpha=1$, (\ref{Zeqn}) can be solved to give 
\begin{equation}
\label{ }
f_n=Nz^2U\left (\hat{\zeta}-a_n,2,\frac{z^2}{R^{\prime\;2}}\right ),\quad\mbox{where}\quad\hat{\zeta}=1+\frac{g^2h_0^2}{8R}\quad\mbox{and}\quad a_n=\frac{m_n^2R^{\prime\;2}}{4}.
\end{equation}
Once again imposing Neumann boundary conditions on the UV brane gives the condition
\begin{displaymath}
2(1-\hat{\zeta}+a_n)zN\left ((a_n-\hat{\zeta})U\left (1+\hat{\zeta}-a_n,2,\Omega^{-2}\right )+U\left (\hat{\zeta}-a_n,2,\Omega^{-2}\right )\right )=0,
\end{displaymath} 
which in turn implies $\hat{\zeta}=1+\frac{M_{W/Z}^2R^{\prime\;2}}{4}$ and
\begin{displaymath}
f_0=Nz^2U\left (1,2,\frac{z^2}{R^{\prime\;2}}\right )=N.
\end{displaymath}
Hence in the case of a linear VEV ($\alpha=1$), the zero mode profile of the W and Z gauge fields are flat. This can alternatively be seen by noting that $\partial_5f_0=0$ satisfies (\ref{Zeqn}) when $g^2h_0^2=2m_0^2RR^{\prime\;2}$. In models with warped extra dimensions, the deformation of the W and Z zero mode gives rise to a number of significant constraints, for example corrections to the $Z\bar{b}b$ vertex, rare lepton decays and corrections to electroweak observables, in particular the $S$ parameter. One would anticipate that a flat $W/Z$ profile would lead to a significant suppression of such constraints, although one would expect that it would also be difficult to generate the fermion mass hierarchy with a linear VEV. 

\subsection{Convergence of Four Fermion Operators}\label{sect:converge}

We are now in a position to test whether or not the coefficients of the four fermion operators, of interest to flavour physics, are divergent (at tree level) with respect to summing over KK number. As we shall see, in section \ref{quarksector}, these coefficients are determined by the integral
\begin{equation}
\label{ ILLLL}
\mathcal{I}=\int_R^{\infty}dz\int_R^{\infty}dz^{\prime}\; f_{L/R}^i(z)f_{L/R}^i(z)G_p(u,v)f_{L/R}^j(z^{\prime})f_{L/R}^j(z^{\prime}).
\end{equation} 
Typically this integral cannot be done analytically, although one can make a small momentum (i.e. small $a$) approximation\footnote{Here we have used that $M(1,2,x)=\frac{e^x-1}{x}$ and Taylor expanded\newline $x^2U\left (1-a,2,\frac{x^2}{R^{\prime\;2}}\right )\approx \frac{R^{\prime\;2}}{\Gamma(1-a)}\left (1+\frac{ax^2}{R^{\prime\;2}}\left [1-2\gamma-\Psi(1-a)-\ln (\frac{x^2}{R^{\prime\;2}})+\frac{x^2}{R^{\prime\;2}}\left (\frac{5-2\Psi(2-a)-4\gamma-2\ln(\frac{x^2}{R^{\prime\;2}}}{4}\right )\right ]+\mathcal{O}\left (\frac{a^2x^4}{R^{\prime\;4}}\right )\right )\approx\frac{R^{\prime\;2}}{\Gamma(1-a)}\left (1+\frac{ax^2}{R^{\prime\;2}}\right )+\mathcal{O}(a^2)$, where $\Psi$ is a digamma function and $\gamma$ is the Euler constant.} 
\begin{equation}
\label{ }
G_p(u,v)\approx\frac{R^{\prime\;2}}{2R}\left (1-\exp\left (\frac{u^2}{R^{\prime\; 2}}\right )-\frac{B(a)}{A(a)}\right )\left (1+\frac{av^2}{R^{\prime\;2}}+\mathcal{O}(a^2)\right ).
\end{equation}
This allows (\ref{ ILLLL}) to be approximated as
\begin{eqnarray}
\mathcal{I}\approx N^{2}_iN^{2}_j\frac{R^{\prime\;2}}{4R}\Bigg [\int_R^{\infty}dz z^{-2c_0^i}\exp\left (\frac{-c_1^iz^2}{R^{\prime\;2}}\right )\left (1-\exp\left (\frac{z^2}{R^{\prime\; 2}}\right )-\frac{B}{A}\right )\Bigg (\left (\frac{c_1^j}{R^{\prime\;2}}\right )^{c_0^j-\frac{1}{2}}\Gamma\left (\frac{1}{2}-c_0^j,\frac{z^2c_1^j}{R^{\prime\;2}}\right ) \nonumber\\
+\frac{a}{R^{\prime\;2}}\left (\frac{c_1^j}{R^{\prime\;2}}\right )^{c_0^j-\frac{3}{2}}\Gamma\left (\frac{3}{2}-c_0^j,\frac{z^2c_1^j}{R^{\prime\;2}}\right )\Bigg )+\int_R^{\infty}dz^{\prime}\left ((z\rightarrow z^{\prime}),\;(i\leftrightarrow j) \right )\Bigg ].
\end{eqnarray} 
The point is that this can be approximately evaluated using
\begin{displaymath}
\int_0^{\infty}dx\;x^{\mu-1}e^{-\beta x}\Gamma(\nu,\alpha x)=\frac{\alpha^\nu\Gamma(\mu+\nu)}{\mu(\alpha+\beta)^{\mu+\nu}}{}_2F_1\left (1,\mu+\nu;\mu+1;\frac{\beta}{\alpha+\beta}\right ),
\end{displaymath}
 but only when $c_1^i+c_1^j-1>0$ and $c_0^i,c_0^j<\frac{1}{2}$ (i.e. only when the fermions are localised towards the IR) \cite{Gradshteyn+Ryzhik}. Alternatively one can check when the integrand blows up. Making a large $x$ expansion of the Kummer functions gives 
\begin{eqnarray}
\mathcal{I}\approx\frac{N_i^2\Gamma(1-a)N_j^2}{2ARR^{\prime\;2}}\Bigg [\int_R^\infty dz \int_{z^{\prime}=z}^\infty dz^{\prime} z^{2-2c_0^i}\exp\left (\frac{-c_1^iz^2}{R^{\prime\;2}}\right )\Bigg [\frac{A}{\Gamma(1-a)}\exp\left (\frac{z^2}{R^{\prime\;2}}\right )\left (\frac{z^2}{R^{\prime\;2}}\right )^{-1-a}\nonumber \\
+B\left (\frac{z^2}{R^{\prime\;2}}\right )^{a-1}\Bigg ]\left (\frac{z^{\prime\;2}}{R^{\prime\;2}}\right )^{a-1}z^{\prime\;2-2c_0^j}\exp\left (\frac{-c_1^jz^{\prime\; 2}}{R^{\prime\;2}}\right )+\int_R^\infty dz^{\prime} \int_{z=z^{\prime}}^\infty dz \left ((z\leftrightarrow z^{\prime}),\;(i\leftrightarrow j) \right )\Bigg ].
\end{eqnarray}
Clearly, for large $z$ the integrand will be dominated by the exponential and so will go to zero only when, once again,
\begin{equation}
\label{divergenceCondition}
c_1^i+c_1^j-1>0.
\end{equation} 
It is straightforward to check this empirically, for the exact integrand, and it is found to hold for every case checked. So, although we have not been able to prove it explicitly, we strongly suspect that the coefficients of four fermion operators will only be convergent when (\ref{divergenceCondition}) is satisfied. That is to say, the four fermion operator coefficients will be divergent, at leading order, if the fermions bulk mass term is not sufficiently $z$-dependent, i.e. the $c_1$ parameter is too small. 

\begin{table}
  \centering 
  \begin{tabular}{|c||cc|cc|cc||cc|cc|cc|}
\hline
% after \\ : \hline or \cline{col1-col2} \cline{col3-col4} ...
   &\multicolumn{6}{|c|}{$n=1$}&\multicolumn{6}{|c|}{$n=2$}  \\
   \hline
  &\multicolumn{2}{|c|}{$c_1=0.5$}&\multicolumn{2}{|c|}{$c_1=1$}&\multicolumn{2}{|c|}{$c_1=1.5$}&\multicolumn{2}{|c|}{$c_1=0.5$}&\multicolumn{2}{|c|}{$c_1=1$}&\multicolumn{2}{|c|}{$c_1=1.5$}  \\
  \hline
  $c_0$ &$\frac{m_n}{M_{\rm{KK}}}$&$\frac{Y_{\rm{eff}}}{\sqrt{2}\lambda}$&$\frac{m_n}{M_{\rm{KK}}}$&$\frac{Y_{\rm{eff}}}{\sqrt{2}\lambda}$&$\frac{m_n}{M_{\rm{KK}}}$&$\frac{Y_{\rm{eff}}}{\sqrt{2}\lambda}$&$\frac{m_n}{M_{\rm{KK}}}$&$\frac{Y_{\rm{eff}}}{\sqrt{2}\lambda}$&$\frac{m_n}{M_{\rm{KK}}}$&$\frac{Y_{\rm{eff}}}{\sqrt{2}\lambda}$&$\frac{m_n}{M_{\rm{KK}}}$&$\frac{Y_{\rm{eff}}}{\sqrt{2}\lambda}$ \\
   \hline\hline
   0.3&1.41&1.90&2.00&1.34&2.45&1.10&2.00&2.62&2.83&1.85&3.46&1.51 \\
   0.4&1.41&1.88&2.00&1.33&2.45&1.09&2.00&2.60&2.83&1.84&3.46&1.50 \\
   0.51&1.43&1.89&2.02&1.34&2.47&1.09&2.01&2.59&2.84&1.83&3.48&1.50 \\
   0.6&1.48&1.93&2.10&1.37&2.57&1.12&2.05&2.62&2.90&1.85&3.55&1.51 \\
   0.7&1.55&1.98&2.19&1.40&2.68&1.14&2.10&2.66&2.97&1.88&3.63&1.53 \\
\hline
\end{tabular}
  \caption{The masses and couplings of the first two KK fermion modes. Included is the effective Yukawa coupling $Y_{\rm{eff}}^{(n,n)}$ for fermion fields that have a zero mode. $M_{\rm{KK}}\equiv\frac{1}{R^{\prime}}=1$ TeV, $\Omega=10^{15}$.  }\label{Tab:FermMasses}
\end{table}

If one interprets the $c_1$ values as arising from the Yukawa couplings then one would anticipate that $c_1\sim h_0 \lambda_{ij} $. With $h_0\sim (250\; \rm{GeV})/ M_{\rm{KK}}$ then naively one would expect that quite large Yukawa couplings ($ \lambda_{ij} $) are needed in order to arrive at order one $c_1$ values. However, as was discussed in \cite{Csaki:2008zd, Casagrande:2008hr}, if the Yukawa couplings are too large one loses perturbative control of the theory. One can use naive dimensional analysis to estimate an upper bound, on the Yukawa couplings, from ensuring perturbative control over the one loop correction to the fermion masses. In particular one requires
\begin{displaymath}
\frac{|Y_{\rm{eff}}|^2}{16\pi^2}\frac{\Lambda^2}{m_n^2}<1,
\end{displaymath} 
where $Y_{\rm{eff}}$ is the effective coupling between the KK fermions and the Higgs. If one requires that the theory is perturbative up to at least the second KK mode ($\Lambda\sim 2m_n$) then this requires that $Y_{\rm{eff}}\lesssim 2\pi$. Before one can compute $Y_{\rm{eff}}$, one must specify the details of the Higgs sector, in particular the bulk and brane potentials. However if we assume that the Higgs VEV is dominated by the zero mode and so approximate the Higgs zero mode profile by $f_H(z)\approx Nh(z)$, such that $\int_R^{\infty}e^{-\Phi}\frac{R^3}{z^3}f_H^2=1$ and hence $N\approx \frac{\sqrt{2}R^{\prime}}{h_0}$, this would imply that
\begin{equation}
\label{ }
\frac{Y_{\rm{eff}}^{(n,m)}}{\sqrt{2}}\approx \frac{Y}{\sqrt{2}}\int_R^\infty dz\;\frac{R}{z}f_Hf_L^{(n)}f_R^{(m)}\approx \lambda\int_R^\infty dz\;\frac{z}{R^{\prime}}f_L^{(n)}f_R^{(m)},
\end{equation}
where the fermion profiles are given in (\ref{KKfermionModes}). These effective couplings are shown in table \ref{Tab:FermMasses}. Hence, as in the RS model, if $c_1\sim\mathcal{O}(1)$ then one would anticipate losing perturbative control of the theory, at a scale lower than the KK scale, when $|\lambda_{ij}|\gtrsim 3$. Unfortunately it is difficult to make any concrete statements, about what scale one loses perturbative control of the theory, since we do not know the exact relation between the Yukawa couplings and the $c_1$ parameter. It is also possible that the $z$-dependent mass arises through some alternative physics other than the Higgs. Due to these uncertainties in the model here we shall not investigate potentially interesting effects arising from large or small $c_1$ values. Rather here we shall consider $c_1$ values for which the results are reasonably independent of the specific $c_1$ value, in particular $c_1=0.5,\; 1,\;1.5$.       

  \begin{table}[h!]
  \centering 
  \begin{tabular}{|c|c||c|c|c|c|}
  \hline
$c_0^i$ & $c_0^j$ & Propagator & 5 KK modes & 10 KK modes & 20 KK modes \\
\hline
\hline
   $ 0.3 $& $0.4$ & $1.7550\times 10^{-8}$ & $1.6729\times 10^{-8}$ & $1.7158\times 10^{-8}$ & $1.7325\times 10^{-8} $\\
     $ 0.3 $& $0.65$ & $-1.8710\times 10^{-8}$ & $-1.7570\times 10^{-8}$ & $-1.8202\times 10^{-8}$ & $-1.8471\times 10^{-8} $\\
  $ 0.6 $& $0.65$ & $6.8463\times 10^{-9}$ & $6.1236\times 10^{-9}$ & $6.4615\times 10^{-9}$ & $6.6265\times 10^{-9} $ \\
\hline
\end{tabular}
\caption{In the propagator column is the quantity $\frac{1}{p^2}-\frac{\mathcal{I}}{f_0^2}$, while in the remaining columns is the quantity $-\frac{1}{f_0^2}\sum_{n=1}^{5, 10, 20}\frac{g_n^ig_n^j}{p^2-m_n^2}$. Here $\Omega=10^2$, $c_1=1$, $p=10$ GeV and $M_{\rm{KK}}\equiv\frac{1}{R^{\prime}}=1$ TeV.} \label{PropCoverge} 
\end{table} 

 \begin{figure}[h!]
    \begin{center}
        \subfigure[]{%
           \label{fig:FermCoupl}
           \includegraphics[width=0.73\textwidth]{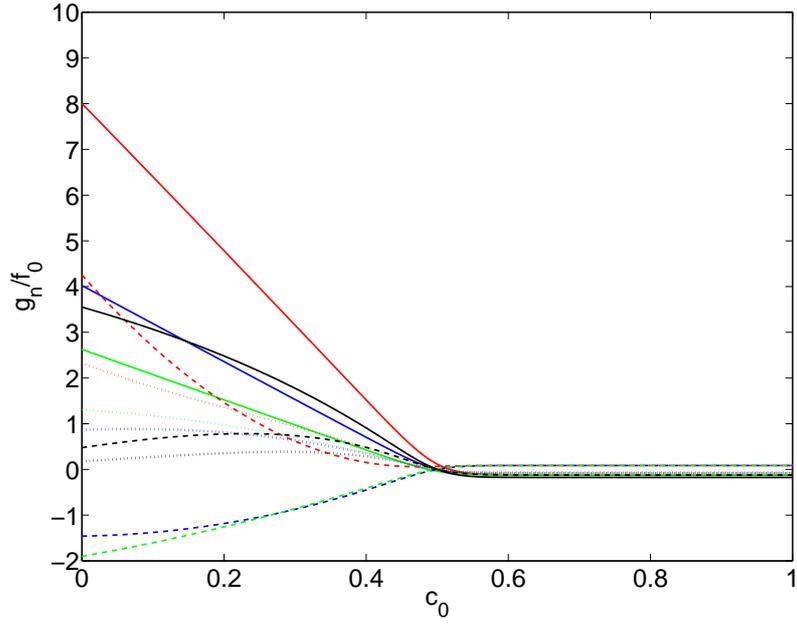}
        }
        \subfigure[]{%
           \label{fig:UVFermCoupl}
           \includegraphics[width=0.73\textwidth]{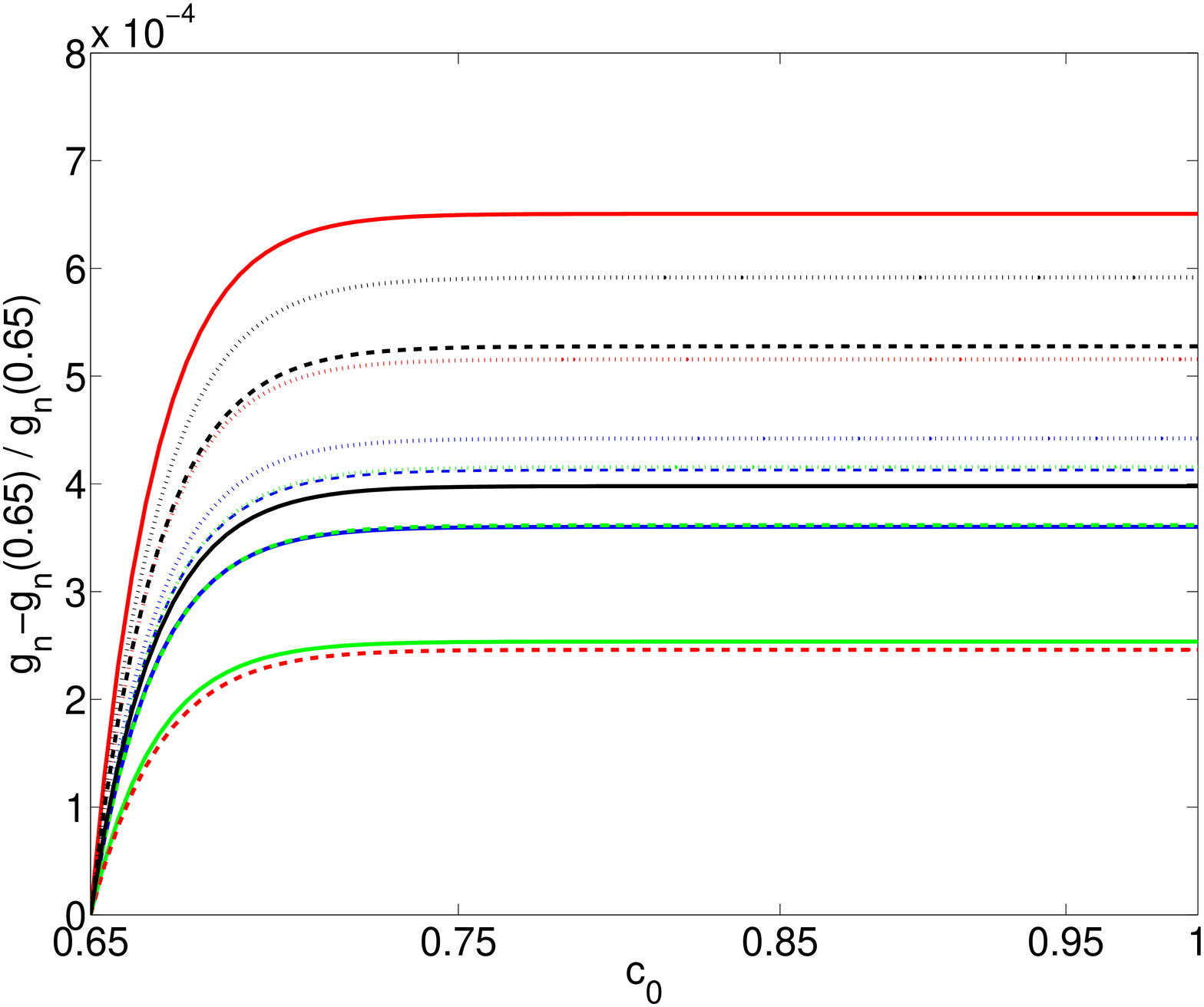}
        }
    \end{center}
    \caption{The relative coupling between the fermion zero modes and the first (solid lines), second (dashed lines) and third (dots) gauge KK modes for the RS model (black) and the SW model with $c_1=0.5$ (red), $c_1=1$ (blue) and $c_1=1.5$ (green). Here $\Omega=10^{15}$ and $M_{\rm{KK}}\equiv \frac{1}{R^{\prime}}=1\;\rm{TeV}$. The lower graph focuses on the universality of the couplings of fermions localised towards the UV brane.} \label{fermCoupl}
\end{figure}

In practice, for large warp factors, the integral (\ref{ ILLLL}) is difficult to do even numerically. In order to carry out the scans over parameter space required to study flavour, it is easier to work with the individual KK modes. Eq. (\ref{ ILLLL}) can be equated to
\begin{equation}
\label{ }
\mathcal{I}=\sum_{n=0}^\infty\frac{g_n^ig_n^j}{p^2-m_n^2},\quad\mbox{where}\quad g_n^i=\int_R^{\infty}dz f_{L/R}^i(z)f_n(z)f_{L/R}^i(z),
\end{equation} 
 and $f_n$ is the gauge profile (\ref{Zprofile} with $\zeta=1$). The question then remains, how many KK modes should be summed over. In table \ref{PropCoverge} we compare this convergence for a low warp factor. Note, at momenta much lower than the KK scale, the propagator is dominated by the zero mode which we have subtracted off. We observe a reasonably good convergence. In practice we sum over the first 15 KK modes and hence would expect an error of the order of a few percent. 
   
Before moving on to look at the results it is worth briefly looking at the relative gauge fermion couplings plotted in figure \ref{fig:FermCoupl}. The only reason why a convergence in (\ref{ ILLLL}) is possible is because the higher KK modes are increasingly weakly coupled to the fermion zero modes. One can also see that the couplings of the RS model do appear to fall away more rapidly than those of the SW model, although it is not really possible to say much when considering just 3 modes. 

The scale of constraints from flavour physics is partly determined by the difference or non-universality of the coupling of different flavours, particularly when they are localised towards the UV brane ($c_0>\frac{1}{2}$). This has been plotted in figure \ref{fig:UVFermCoupl} for an arbitrary bulk mass parameter of $c_0=0.65$. Firstly one can see the source of the so called RS-GIM mechanism since, when the the fermions are sitting towards the UV brane, the gauge fermion coupling is approximately universal. One can also see, from figure \ref{fig:UVFermCoupl}, that the SW model has an equivalent level of universality to that of the RS model. Critically, one can also see from figure \ref{fig:FermCoupl}, that when the fermions are localised towards the UV, differences in the couplings are dominated by differences in the $c_0$ bulk mass term and not the $c_1$ term. Hence here assuming universal $c_1$ parameters does not significantly change the results. However as one localises the fermions further and further towards the IR then one can see that the couplings becomes increasingly sensitive to the $c_1$ parameter.

%%%%%%%%%%%%%%%%%%%%%%%%%%%%%%%%%%%%%%%%%%%%%%%%%%%%
\section{The Lepton Sector}\label{sect:Lept}

Before considering the quark sector we shall first look at the tree level decays involving just the charged leptons. The advantage to this is that one need only fit to the lepton masses and this will allow us to demonstrate the central physics a little more clearly. In light of the current experimental status of the PMNS matrix one is inclined to favour configurations with large charged lepton mixings (i.e. closely spaced $c_0$ parameters). However here we will consider scenarios with both large and small mixings.  

We shall also consider only the zero modes of the fermions. One may be concerned that, when the full mass matrix $MM^\dag$ is diagonalised, the mixing of the zero modes would receive contributions from the terms that are off diagonal with respect to KK number. In the RS model these off diagonal terms are partially suppressed by the orthogonality of the fermion profiles (\ref{orthogferm}) \cite{Huber:2003tu}. One would expect a similar effect  being present in the SW model although it would be partially reduced due to the presence of the Higgs profile. As in the case of the RS model, one should be particularly cautious, about neglecting the contribution from the fermion KK modes, when one has matching $c_0$ values. Since this gives rise to particularly universal couplings which can, if the KK modes are not considered, exaggerate the suppression of FCNC's.          

\begin{figure}[ht!]
\begin{center}
\includegraphics[width=0.6\textwidth]{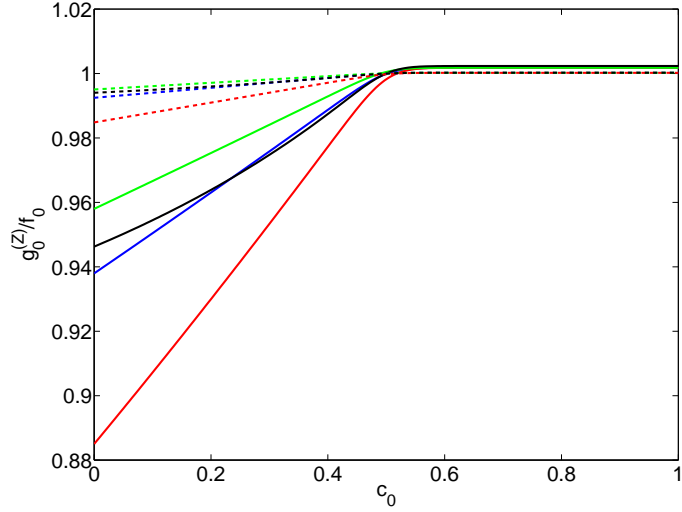}
\caption{The coupling of the Z zero mode for the RS model (black) and the SW model with $c_1=0.5$ (red), $c_1=1$ (blue) and $c_1=1.5$ (green). Here $\Omega=10^{15}$ while $M_{\rm{KK}}= 1$ TeV (solid lines) and $M_{\rm{KK}}= 3$ TeV (dashed lines). }
\label{fig:Zcoupling}
\end{center}
\end{figure}

The process we will consider here is the tree level decay $l_j\rightarrow l_il_i\bar{l}_i$ which is given by \cite{Langacker:2000ju}
\begin{displaymath}
\Gamma(l_j\rightarrow l_il_i\bar{l}_i)=\frac{G_F^2m_{l_j}^5}{48\pi^3}\left (2|\mathcal{C}^L_{ij}|^2+2|\mathcal{C}^R_{ij}|^2+ |\mathcal{D}^L_{ij}|^2+|\mathcal{D}^R_{ij}|^2\right ),
\end{displaymath}
where 
\begin{displaymath}
\mathcal{C}_{ij}^{L/R}=\sum_n\frac{M_Z^2}{m_n^2}\left (\mathcal{B}_{(n)}^{L/R}\right )_{ij}\left (\mathcal{B}_{(n)}^{L/R}\right )_{ii}\quad\mbox{and}\quad \mathcal{D}_{ij}^{L/R}=\sum_n\frac{M_Z^2}{m_n^2}\left (\mathcal{B}_{(n)}^{L/R}\right )_{ij}\left (\mathcal{B}_{(n)}^{R/L}\right )_{ii} ,
\end{displaymath}
and
\begin{displaymath}
\mathcal{B}_{(n)}^{L/R}=\frac{1}{f_0}U_{L/R}\;g_n\;U_{L/R}^\dag.
\end{displaymath}
$U_{L,R}$ are the unitary matrices that diagonalise the fermion mass matrix. This process receives contributions from both the exchange of KK photons and KK Z bosons but in practice we find it to be completely dominated by the deformation of the Z zero mode plotted in figure \ref{fig:Zcoupling}. We do not include the contribution from the Higgs or KK modes of the Higgs. The current experimental bounds on these processes are \cite{Nakamura:2010zzi, Hayasaka:2010np} 
\begin{eqnarray}
\mbox{Br}(\mu^{-}\rightarrow e^-e^+e^-)<1.0\times10^{-12}, \nonumber\\ 
  \mbox{Br}(\tau\rightarrow \mu^-\mu^+\mu^-)<2.1\times10^{-8}, \nonumber\\    
  \mbox{Br}(\tau^{-}\rightarrow e^-e^+e^-)<2.7\times10^{-8}.\nonumber
\end{eqnarray}

\subsection{Numerical Analysis and Results}
This slightly simplified study essentially serves three purposes. Firstly, it allows us to look at the central physics for a relatively simple example. Secondly, it allows us to compare the results from the present model, with a $z$-dependent mass (\ref{bulkmass}), with a model in which one uses the Yukawa couplings to get a discrete fermion spectrum \cite{Gherghetta:2009qs, Atkins:2010cc}. Thirdly, it allows us to test the validity of the assumption of a universal $c_1$ value. This is rather crudely tested by assuming a universal $c_1$ value and then looking at the results for any $c_1$ dependence. By considering just these three decays, it is reasonable to just fit to the three charged lepton masses. However, one still has a sizeable number of input parameters. Although we assume a universal $c_1$ value, we still allow for anarchic Yukawa couplings $\lambda$ in (\ref{SWMasses}). We shall also assume real, flavour-diagonal $c_0$ values.

\begin{figure}[ht!]
    \begin{center}
        \subfigure[]{%
           \label{fig:ueee}
           \includegraphics[width=0.85\textwidth]{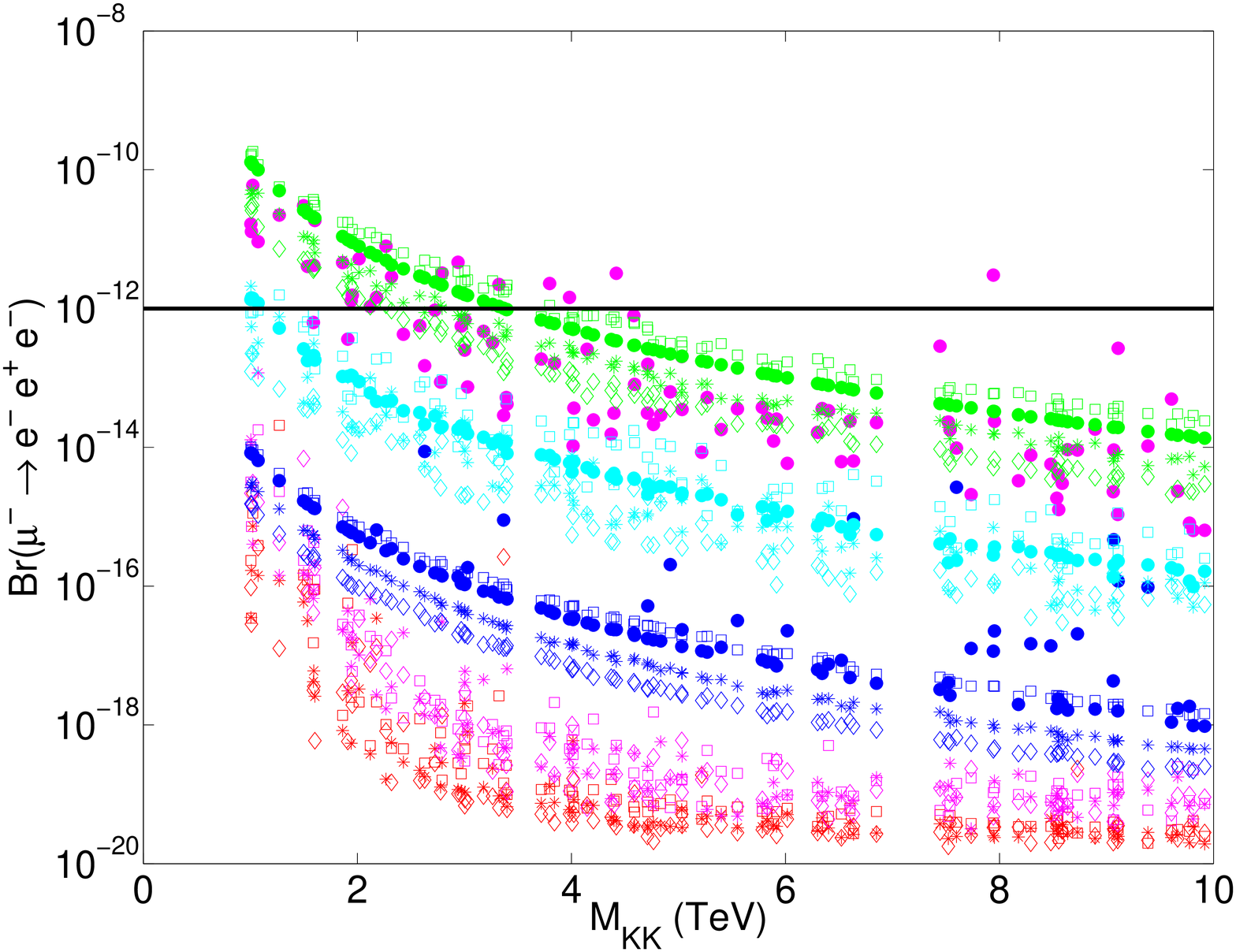}
        }\\
        \subfigure[]{%
           \label{fig:fig:teee}
           \includegraphics[width=0.48\textwidth]{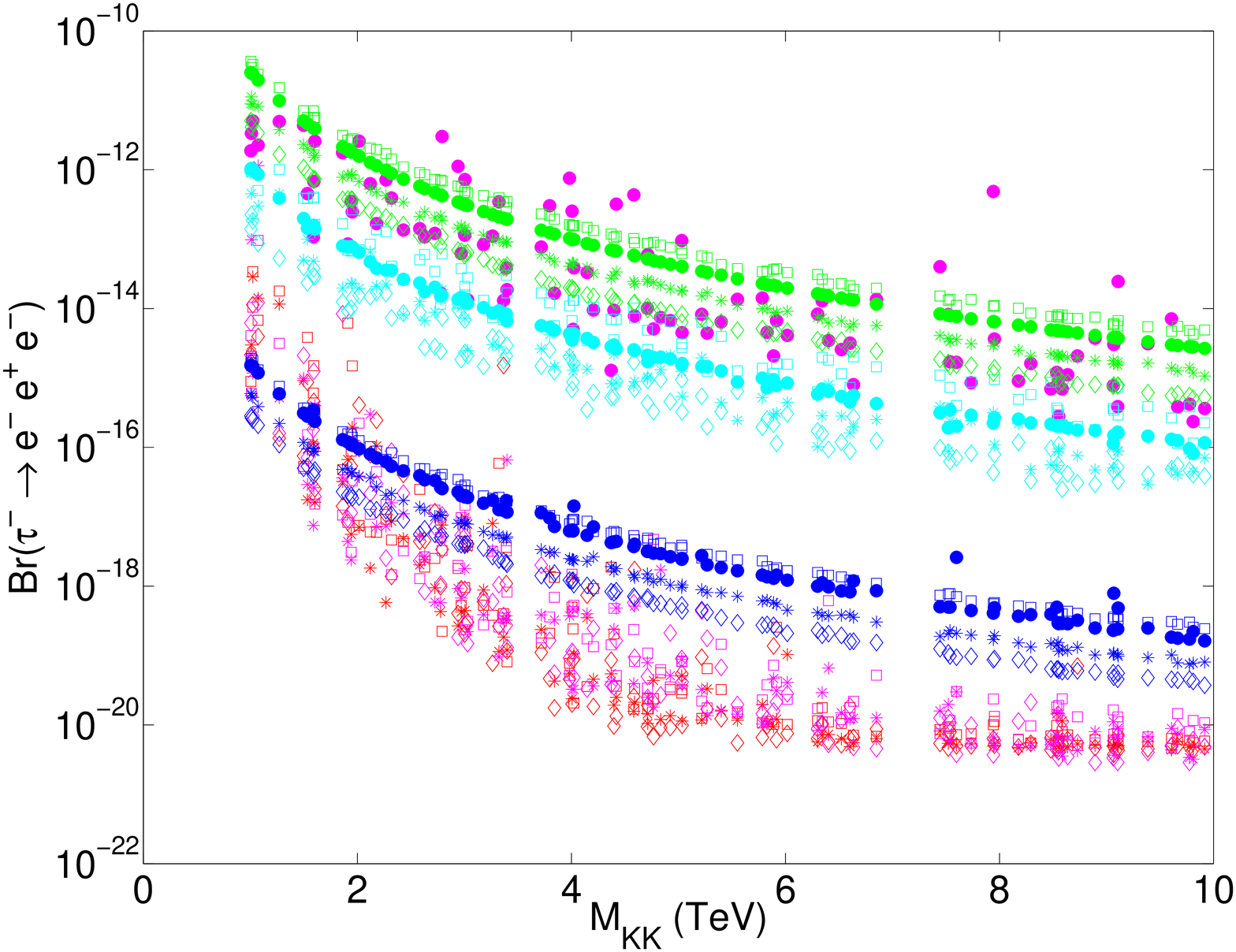}
        }
        \subfigure[]{%
           \label{fig:teee}
           \includegraphics[width=0.48\textwidth, height=0.355\textwidth]{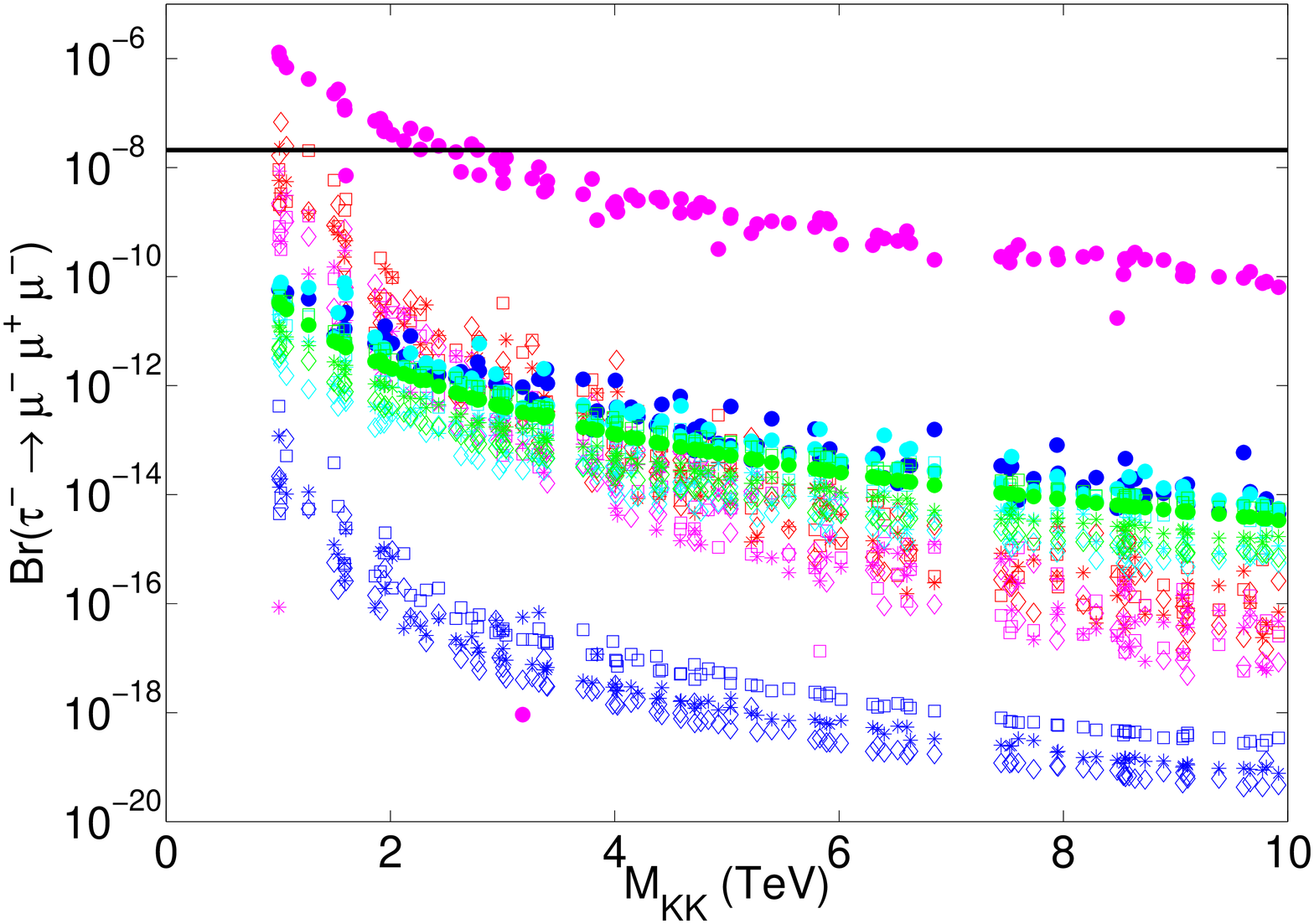}
        }
    \end{center}
    \caption{Branching ratios for rare lepton decays in the RS model (solid dots) and the SW model with $c_1=0.5$ (square), $c_1=1$ (star) and $c_1=1.5$ (diamond). The five configurations considered are given in (\ref{leptConfig}), i.e. (A) in red, (B) in magenta, (C) in blue, (D) in cyan and (E) in green. Note the RS configuration (A) points have not been plotted due to the difficulty in obtaining a reasonable fit to the masses. The black lines indicate the experimental bound although the branching ratios for $\tau^-\rightarrow e^-e^+e^-$ lie well below the experimental bound. $\Omega=10^{15}$.} \label{fig:leptonDecay}
\end{figure}

Here we take the $c_L\;(c_0^{(L)})$ as input parameters and randomly generate ten $3\times 3$ complex Yukawa couplings, $|\lambda_{ij}|\in[1,3]$, allowing the $c_R$ values to be solved for by fitting to the lepton masses \cite{Nakamura:2010zzi}.
\begin{displaymath}
m_e=0.511\;\mbox{MeV}\quad m_\mu=106\;\mbox{MeV}\quad m_\tau=1780\;\mbox{MeV}. 
\end{displaymath}     
We then proceed to generate a further 10,000 Yukawa couplings and for each one compute the branching ratios and lepton masses. Inevitably most of these configurations will not give the correct masses, and so we take the 100 configurations which give the most accurate masses. From these 100 configurations we plot the average of the branching ratios in figure \ref{fig:leptonDecay}. We then repeat this for a hundred random KK scales $M_{\rm{KK}}\equiv \frac{1}{R^{\prime}}\in [1,10]\;$TeV, three $c_1$ values and five $c_L$ values;
\begin{eqnarray}
(A)\quad c_L=[0.710,\;0.700,\;0.690]\quad\quad\quad(B)\quad c_L=[0.750,\;0.700,\;0.650] \nonumber\\
(C)\quad c_L=[0.601,\;0.600,\;0.599]\quad\quad\quad(D)\quad c_L=[0.650,\;0.600,\;0.550] \nonumber\\
(E)\quad c_L=[0.560,\; 0.550,\; 0.540].\label{leptConfig}
\end{eqnarray} 
We then proceed to compute the branching ratios for the RS model using exactly the the same method. In the literature there always appears to be a slight debate over what should be referred to as the KK scale. Here we define the KK scale to be $\frac{1}{R^{\prime}}$ for the SW model and $\frac{k}{\Omega}$ for the RS model. However in the interests of comparison, when plotting the RS points, we rescale $M_{\rm{KK}}$ by a factor of $\frac{2.0}{2.45}$ such that the mass of the first KK gauge mode will, for both models, be about two times $M_{\rm{KK}}$. The results are plotted in figure \ref{fig:leptonDecay}.     

Firstly, it should be noted that the (C) configuration, with $c_1\approx 1$, is in good agreement with the equivalent calculation done without the $z$ dependent mass term \cite{Atkins:2010cc}. Secondly, as one would expect, there is an increasing $c_1$ dependence as one moves the $f_L$ profile towards the IR, although this dependence is still small compared to changes in the $c_0$ parameters, e.g. going from configuration (C) to (D). None the less we find that this dependence is negligible for configurations (A) and (B). Bearing in mind that these configurations typically have $c_0^R$ values sitting further towards the UV this suggests that the assumption of universal $c_1$ values is good when $c_0^L\gtrsim 0.6\;(\mbox{ or }\;c_0^R\lesssim-0.6)$.    

These figures also highlight the implications of the reduced $c_0$ dependence in the range of fermion masses (i.e. the gradient in figure \ref{fermmass}). One can clearly see that as one localises the left-handed fermions closer and closer towards the UV brane (i.e. larger $c_L$ values), the scale of the branching ratios is reduced. In order to maintain the correct masses, the corresponding right-handed fermions must sitter closer towards the IR. Sooner or later this leads to a problem via either a large value of $\mbox{Br}(\tau^-\rightarrow \mu^-\mu^+\mu^-)$ or even a difficulty obtaining the correct masses. However, due to this reduced gradient the corresponding $c_0^R$ are not as extreme in the SW model as in the RS model. Hence the RS model has problematic points in parameter space before the SW model. For example a $\mbox{Br}(\mu^-\rightarrow e^-e^+e^-)<10^{-18}$ is difficult to achieve in the RS model but not in the SW model. In other words the reduced $c_0$ dependence, in the SW model, gives rise to a larger phenomenologically viable parameter space than in the RS model.  

In a more complete study, that included processes such as $\mu\rightarrow e\gamma$, it would be necessary to include both neutrino masses as well as mixings. This would restrict the range of possible configurations (\ref{leptConfig}) but one would anticipate that the basic results still hold. 

%%%%%%%%%%%%%%%%%%%%%%%%%%%%%%%%%%%%%%%%%%%%%%%%%%%%
\section{The Quark Sector}\label{quarksector}
As mentioned in the introduction, some of the most stringent
constraints on the RS model come from FCNC's in the kaon sector \cite{
  Csaki:2008zd, Blanke:2008zb, Bauer:2009cf, Gedalia:2009ws,
  Agashe:2008uz}, so we wish to see how they are affected in
the SW model. As in
\cite{Csaki:2008zd, Bauer:2009cf, Chang:2008vx}, we integrate out the
the non-zero gluon and photon KK modes, as well as
all weak gauge boson modes, at tree level, arriving at the 4D
effective Hamiltonian
\cite{Bagger:1997gg}
\begin{equation}
\label{weakham}
\mathcal{H}_{\mbox{eff}}^{\triangle
  S=2}=\sum_{i=1}^5C_iQ_i^{sd}+\sum_{i=1}^3\tilde{C}_i\tilde{Q}_i^{sd} .
\end{equation}
Here
\begin{eqnarray}
Q_1^{sd}=(\bar{d}_L\gamma^\mu s_L)(\bar{d}_L\gamma_\mu s_L)\qquad \tilde{Q}_1^{sd}=(\bar{d}_R\gamma^\mu s_R)(\bar{d}_R\gamma_\mu s_R)\nonumber\\
Q_2^{sd}=(\bar{d}_R s_L)(\bar{d}_R s_L)\hspace{1.45cm} \tilde{Q}_2^{sd}=(\bar{d}_L s_R)(\bar{d}_L s_R)\hspace{0.8cm}\nonumber\\
Q_3^{sd}=(\bar{d}^\alpha_R s^\beta_L)(\bar{d}^\beta_R s^\alpha_L)\hspace{1.45cm} \tilde{Q}_3^{sd}=(\bar{d}^\alpha_L s^\beta_R)(\bar{d}^\beta_L s^\alpha_R)\hspace{0.8cm}\nonumber\\
Q_4^{sd}=(\bar{d}_Rs_L)(\bar{d}_Ls_R)\hspace{5.3cm}\nonumber\\
Q_5^{sd}=(\bar{d}^\alpha_Rs^\beta_L)(\bar{d}^\beta_Ls^\alpha_R)\hspace{5.3cm}\nonumber 
\end{eqnarray}
and $\alpha$ and $\beta$ are colour indices. Model-independent bounds
on the Wilson coefficients $C_i, \bar C_i$ have been
given in \cite{Bona:2007vi}, as quoted in Table \ref{tab:utfitbounds}.
\begin{table}
\begin{center}
\begin{tabular}{|cc|cc|}
\hline
  Parameter & $95\%$ allowed range (GeV${}^{-2}$) &Parameter & $95\%$ allowed range (GeV${}^{-2}$)  \\
  \hline
  \hline
 Re $C_1$  & $[-9.6,\;9.6]\;\times 10^{-13} $& Im $C_1$  & $[-4.4,\;2.8]\;\times 10^{-15} $  \\
 Re $C_2$  & $[-1.8,\;1.9]\;\times 10^{-14} $& Im $C_2$  & $[-5.1,\;9.3]\;\times 10^{-17} $  \\
 Re $C_3$  & $[-6.0,\;5.6]\;\times 10^{-14} $& Im $C_3$  & $[-3.1,\;1.7]\;\times 10^{-16} $  \\
 Re $C_4$  & $[-3.6,\;3.6]\;\times 10^{-15} $& Im $C_4$  & $[-1.8,\;0.9]\;\times 10^{-17} $  \\
 Re $C_5$  & $[-1.0,\;1.0]\;\times 10^{-14} $& Im $C_5$  & $[-5.2,\;2.8]\;\times 10^{-17} $  \\
\hline
\end{tabular}
\end{center}
\caption{
\label{tab:utfitbounds}
Allowed ranges for the $\Delta F=2$ Wilson coefficients \cite{Bona:2007vi}.
}
\end{table}
Let us define the integral
\begin{equation}
\label{fourFermInt}
\mathbb{I}_{\psi_k\chi_l\xi_m\sigma_n}^{(A)}=\sum_{i,j=1}^3(U_\psi^\dag)^{ki}(U_\chi)^{il}\left
  [\int_R^\infty dz\int_R^\infty dz^{\prime}f_\psi^i(z)
  f_\chi^i(z)G_p^{(A)}(u,v)f_\xi^j(z^{\prime})f_\sigma^j(z^{\prime})\right
](U_\xi^\dag)^{mj}(U_\sigma)^{jn} ,
\end{equation}  
where $\psi,\chi,\xi,\sigma=L,R$ while $U_{L/R}$ are the unitary matrices that diagonalise the mass matrices  and $i,j,k,l,m,n$ are flavour indices. The Wilson coefficients are then given by
\begin{eqnarray}
C_1
&=&\frac{g_s^2}{6}\mathbb{I}_{L_dL_sL_dL_s}^{(G)}+\frac{g^2}{2c_w^2}\bigg(\frac{1}{2}-\frac{1}{3}s_w^2\bigg)^2\mathbb{I}_{L_dL_sL_dL_s}^{(Z)}+\frac{e^2}{18}\mathbb{I}_{L_dL_sL_dL_s}^{(A)}
\; ,\\
\tilde{C}_1&=&\frac{g_s^2}{6}\mathbb{I}_{R_dR_sR_dR_s}^{(G)}+\frac{g^2
  s_w^4}{18c_w^2}\mathbb{I}_{R_dR_sR_dR_s}^{(Z)}+\frac{e^2}{18}\mathbb{I}_{R_dR_sR_dR_s}^{(A)}
\; ,\\
C_4&=&-g_s^2\;\mathbb{I}_{L_dL_sR_dR_s}^{(G)} \; ,\\
C_5&=&\frac{g_s^2}{3}\mathbb{I}_{L_dL_sR_dR_s}^{(G)}+\frac{2g^2s_w^2}{3c_w^2}\bigg
(\frac{1}{2}-\frac{1}{3}s_w^2\bigg
)\mathbb{I}_{L_dL_sR_dR_s}^{(Z)}-\frac{2e^2}{9}\mathbb{I}_{L_dL_sR_dR_s}^{(A)}\; ,
\end{eqnarray}
where $s_w^2 = \sin^2 \theta_W$, $c_w^2 = \cos^2 \theta_W$,
$\theta_W$ is the weak mixing angles and $g_s,\; g,\; e$ are
the 5D couplings of the (five-dimensional) gluon, Z and photon fields.
Here we are not interested in electroweak constraints, so we
shall equate $g_s^2f_0^2=4\pi\alpha_s$, $g^2f_0^2=4\pi\alpha/s_w^2$,
and $e^2f_0^2=4\pi\alpha$.

The $K_L-K_S$ mass difference and the indirect CP violation
parameter $\epsilon_K$ follow from the weak
Hamiltonian~(\ref{weakham}). Because the former is more sensitive to
uncertain long-distance effects than the latter and because of the small
experimental value
$(\epsilon_K)_{\rm exp}=(2.228\pm0.011)\times10^{-3}$ \cite{Nakamura:2010zzi},
the stronger constraints are on the (CP-violating) imaginary parts of
the Wilson coefficients (cf Table \ref{tab:utfitbounds}). These stronger
constraints typically translate into the strongest constraints on the
model, as is the case in the RS model (or the MSSM, for that
matter). We expect the same to hold for the SW model.
The new-physics (NP) contributions to $\epsilon_K$ and
$\Delta M_K$ are calculated using \cite{Bauer:2009cf}
\begin{equation}
\label{epsk}
\epsilon_K=\frac{\kappa_\epsilon
  e^{i\varphi_\epsilon}}{\sqrt{2}(\triangle
  m_K)_{\rm{exp}}}\rm{Im}\,\langle
K^0|\mathcal{H}_{\mbox{eff}}^{\triangle S=2}|\bar{K}^0 \rangle\; ,
\end{equation} 
\begin{equation}
\label{ }
\Delta m_K=2\, \rm{Re}\,\langle
K^0|\mathcal{H}_{\mbox{eff}}^{\triangle S=2}|\bar{K}^0 \rangle\;
\end{equation}
where $\varphi_\epsilon=43.51^\circ$ and
$\kappa_\epsilon=0.92$ \cite{Buras:2008nn}. The hadronic matrix elements of the
four-fermion operators are parameterised as
\begin{eqnarray*}
\langle K^0|Q_1^{sd}(\mu)|\bar{K}^0
\rangle&=&\frac{m_Kf_K^2}{3}B_1(\mu) \; , \\
\langle K^0|Q_4^{sd}(\mu)|\bar{K}^0 \rangle&=&\left
  (\frac{m_K}{m_d(2M_{\rm{KK}})+m_s(2M_{\rm{KK}})}\right
)^2\frac{m_Kf_K^2}{4}B_4(\mu) \; ,\\
\langle K^0|Q_5^{sd}(\mu)|\bar{K}^0 \rangle&=&\left
  (\frac{m_K}{m_d(2M_{\rm{KK}})+m_s(2M_{\rm{KK}})}\right
)^2\frac{m_Kf_K^2}{12}B_5(\mu) \; ,
\end{eqnarray*}
where $m_K=497.6$ MeV, $f_K=156.1$ MeV, and we have indicated the
dependence on the renormalisation scale $\mu$. Tremendous effort has
gone into higher-order corrections to the SM calculation of $C_1$,
to the formula (\ref{epsk}) , and 
the determination of the $B_i$-factors, in particular the
SM one, $B_1$.
In fact, some recent studies have hinted that the SM contribution may
be slightly too small to explain the experimental
value \cite{Buras:2009pj,Lunghi:2008aa}. While this is intriguing,
in view of our tree-level analysis and other approximations made, we will
ignore the SM contributions, comparing the new-physics contribution in
this model directly to the experimental values for $\epsilon_K$ and $\Delta M_K$.
We also restrict ourselves to a leading-log analysis, renormalising our
Wilson coefficients at $\mu = \mu_0 = 2 M_{KK}$ and evolving
\cite{Buras:2001ra} (at leading log) the bag factors
given in \cite{Babich:2006bh} up to $\mu_0$.
Representative numerical values are given in Table~\ref{tab:Bfac}.
\begin{table}
\begin{center}
\begin{tabular}{|c|cccc|}
\hline
   & 1 TeV & 3 TeV & 10 TeV & 30 TeV \\
 \hline  
  $B_1$ & 0.407 & 0.395 & 0.384 & 0.374 \\
  $B_4$& 0.938& 0.938 & 0.938 & 0.938\\
   $B_5$& -0.336 & -0.373  & -0.404 &-0.427 \\
\hline
\end{tabular}
\end{center}
\caption{
\label{tab:Bfac}
``Bag'' factors $B_i(\mu)$ for the relevant hadronic matrix elements at
representative renormalisation scales. }
\end{table}

\subsection{Numerical Analysis and Results}\label{sect:NumRes}
Here we wish to investigate two points. Firstly, can the SW model reproduce the correct quark masses and mixing angles and secondly, to what extent are FCNC's suppressed. However, even if one assumes real, flavour-diagonal, bulk mass parameters and universal, order unity, $c_1$ values then one still has 18 complex Yukawas and 9 real $c_0$ parameters to fit. Such a large parameter space gives rise to an under constrained problem or in other words the existing constraints from flavour physics can always be satisfied with sufficient tuning of the free parameters. Hence the relevant question, we wish to address here, is which of the two models requires the least tuning in order to reproduce all existing observables. Ideally one should carry out a full Monte Carlo analysis, although accurately carrying out the integrals is numerically too slow for this to be a computationally viable option. Another possible approach would be to compute the fine tuning parameter (introduced in \cite{Barbieri88}) including all know observables. However this approach would offer no indication as to the `typical' size of a given observable in a given model. 

The approach taken here is to find points in parameter space that give the correct masses, mixing angles and Jarlskog invariant and then proceed to calculate the size of the additional contributions, to $\epsilon_K$ and $\Delta m_K$, from the KK gauge fields. In selecting such points one should be aware of two factors. Firstly, is the point fine tuned, i.e. are the output observables sensitive to small changes in the input parameters. Secondly, is the point a particularly rare point in parameter space. In order to address the second issue, here we endeavour to scan over as wide a range of the parameter space as is computationally viable. For the first point we will compute the fine tuning parameter at each point considered. To be a little more explicit our method is as follows.

\begin{table}
  \flushleft
  \begin{tabular}{|c|c|c|c|}
\hline
% after \\ : \hline or \cline{col1-col2} \cline{col3-col4} ...
   \multicolumn{2}{|c|}{Configuration}&$c_R^u$&$c_R^d$  \\
   \hline\hline
\multirow{4}{*}{(A)}&$c_1=0.5$&$[-0.66\pm0.04,\;-0.47\pm0.12,\;0.46\pm0.10]$& $[-0.63\pm0.01,\;-0.61\pm0.01,\;-0.57\pm0.01]$ \\
&$c_1=1$&$[-0.65\pm0.04,\;-0.45\pm0.11,\;0.47\pm0.07]$& $[-0.62\pm0.01,\;-0.60\pm0.01,\;-0.56\pm0.01]$ \\
&$c_1=1.5$&$[-0.64\pm0.03,\;-0.43\pm0.13,\;0.45\pm0.13]$& $[-0.62\pm0.01,\;-0.59\pm0.01,\;-0.56\pm0.01]$  \\ \cline{2-4}
&RS&$[-0.62\pm0.01,\;-0.44\pm0.05,\;4.63\pm1.98]$& $[-0.60\pm0.01,\;-0.58\pm0.01,\;-0.55\pm0.01]$  \\
\hline\hline
\multirow{4}{*}{(B)}&$c_1=0.5$&$[-0.69\pm0.03,\;-0.52\pm0.03,\;0.38\pm0.18]$& $[-0.66\pm0.01,\;-0.61\pm0.01,\;-0.60\pm0.01]$ \\
&$c_1=1$&$[-0.69\pm0.07,\;-0.46\pm0.14,\;0.46\pm0.13]$& $[-0.65\pm0.01,\;-0.61\pm0.01,\;-0.58\pm0.01]$ \\
&$c_1=1.5$&$[-0.67\pm0.01,\;-0.46\pm0.12,\;0.45\pm0.15]$& $[-0.65\pm0.01,\;-0.60\pm0.01,\;-0.58\pm0.01]$  \\ \cline{2-4}
&RS&$[-0.65\pm0.01,\;-0.48\pm0.04,\;1.09\pm0.67]$& $[-0.63\pm0.01,\;-0.59\pm0.01,\;-0.57\pm0.01]$  \\
\hline\hline
\multirow{4}{*}{(C)}&$c_1=0.5$&$[-0.72\pm0.04,\;-0.56\pm0.01,\;-0.25\pm0.22]$& $[-0.69\pm0.01,\;-0.65\pm0.01,\;-0.62\pm0.01]$ \\
&$c_1=1$&$[-0.71\pm0.01,\;-0.53\pm0.08,\;0.06\pm0.33]$& $[-0.68\pm0.01,\;-0.64\pm0.01,\;-0.61\pm0.01]$ \\
&$c_1=1.5$&$[-0.70\pm0.03,\;-0.52\pm0.06,\;0.30\pm0.32]$& $[-0.68\pm0.01,\;-0.63\pm0.01,\;-0.60\pm0.01]$  \\ \cline{2-4}
&RS&$[-0.68\pm0.01,\;-0.53\pm0.01,\;-0.06\pm0.14]$& $[-0.66\pm0.01,\;-0.62\pm0.01,\;-0.60\pm0.01]$  \\
\hline\hline
\multirow{4}{*}{(D)}&$c_1=0.5$&$[-0.74\pm0.01,\;-0.60\pm0.01,\;-0.37\pm0.08]$& $[-0.72\pm0.01,\;-0.67\pm0.01,\;-0.62\pm0.01]$ \\
&$c_1=1$&$[-0.73\pm0.01,\;-0.58\pm0.01,\;-0.23\pm0.21]$& $[-0.71\pm0.01,\;-0.67\pm0.01,\;-0.61\pm0.01]$ \\
&$c_1=1.5$&$[-0.73\pm0.01,\;-0.56\pm0.06,\;-0.15\pm0.24]$& $[-0.71\pm0.01,\;-0.66\pm0.01,\;-0.61\pm0.01]$  \\ \cline{2-4}
&RS&$[-0.72\pm0.01,\;-0.57\pm0.01,\;-0.29\pm0.06]$& $[-0.69\pm0.01,\;-0.65\pm0.01,\;-0.61\pm0.01]$  \\
\hline\hline
\multirow{4}{*}{(E)}&$c_1=0.5$&$[-0.77\pm0.01,\;-0.63\pm0.01,\;-0.43\pm0.03]$& $[-0.75\pm0.01,\;-0.71\pm0.01,\;-0.63\pm0.01]$ \\
&$c_1=1$&$[-0.76\pm0.01,\;-0.62\pm0.01,\;-0.36\pm0.07]$& $[-0.74\pm0.01,\;-0.71\pm0.01,\;-0.62\pm0.01]$ \\
&$c_1=1.5$&$[-0.76\pm0.01,\;-0.62\pm0.01,\;-0.21\pm0.24]$& $[-0.73\pm0.01,\;-0.70\pm0.01,\;-0.61\pm0.01]$  \\ \cline{2-4}
&RS&$[-0.74\pm0.01,\;-0.61\pm0.01,\;-0.35\pm0.05]$& $[-0.72\pm0.01,\;-0.69\pm0.01,\;-0.61\pm0.01]$  \\
\hline
\end{tabular}
  \caption{Bulk mass parameters ($c_R=c_0^{Ri}$) of the right-handed fermions obtained by fitting to quark masses and mixing angles with the left-handed fermions having the bulk configurations; $(A)\; c_L=[0.72,\;0.64,\;0.52]$, $(B)\; c_L=[0.69,\;0.63,\;0.49] $, $(C)\; c_L=[0.66,\;0.60,\;0.42]$, $(D)\; c_L=[0.63,\;0.57,\;0.34] $, $(E)\; c_L=[0.60,\;0.52,\;0.25]$. Quoted is the mean and standard deviation taken over 50 random points, see text.}\label{Tab:BulkmassParams}
\end{table}

\begin{figure}[t!]
\begin{center}
\vspace{-1cm}
 \subfigure[]{%
           \label{Fig:Epk}
           \vspace{-4cm}
\includegraphics[width=0.76\textwidth]{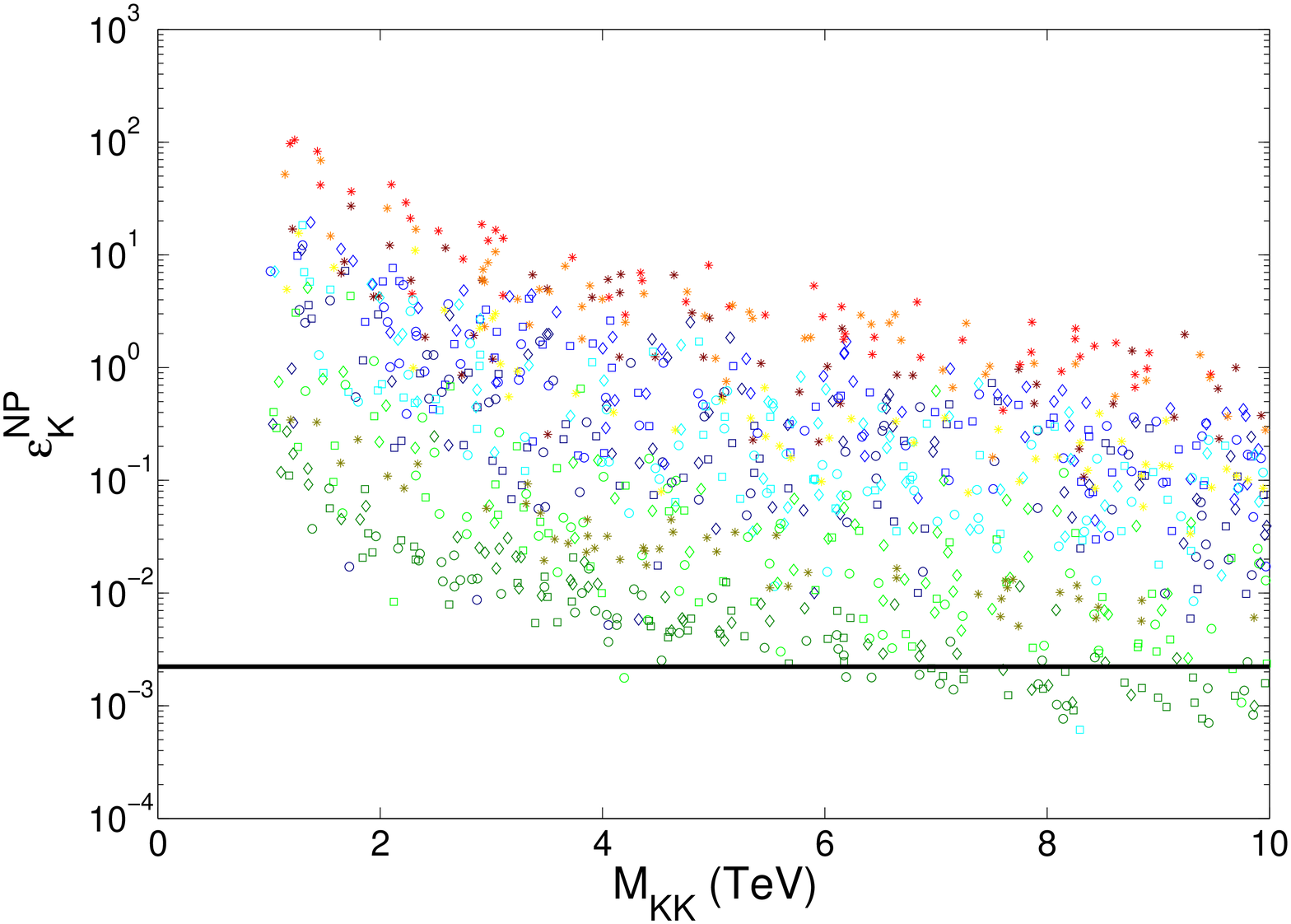}
        }
        \subfigure[]{%
           \label{Fig:MK}
           \vspace{-1cm}
           \includegraphics[width=0.76\textwidth]{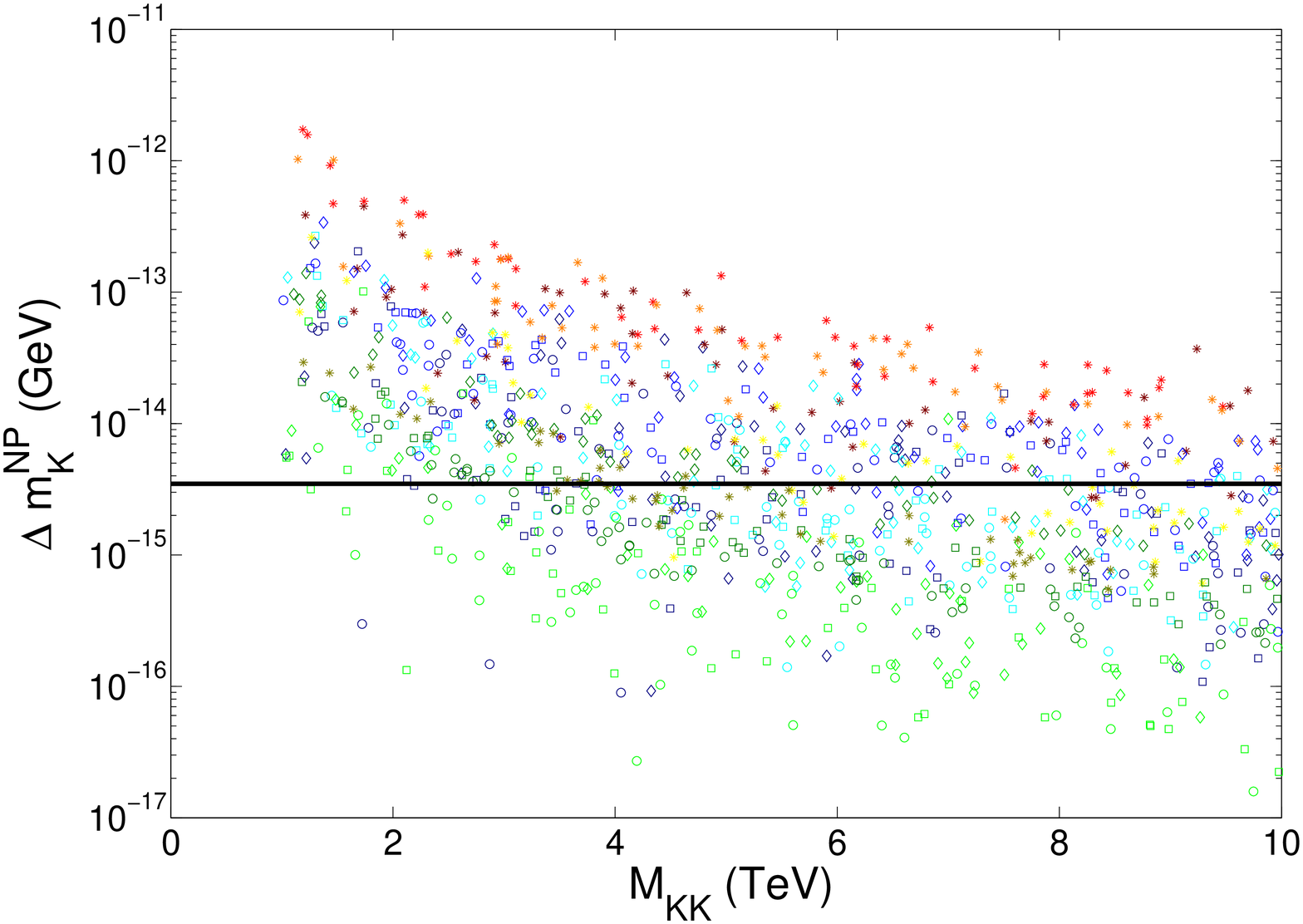}
        }
\caption{The mean values of $\epsilon_K^{\rm{NP}}$ and $\Delta m_K^{\rm{NP}}$ for the RS model (stars) and the SW model with $c_1=1.5$ (circles), $c_1=1$ (squares) and $c_1=0.5$ (diamonds). The $c_L$ values are given in (\ref{SWquarkConfig}). For the SW model configuration (A) is plotted in dark blue, (B) is plotted in light blue, (C) is plotted in cyan, (D) is plotted in light green and (E) is plotted in dark green. While for the RS model (A) is plotted in dark red, (B) is plotted in light red, (C) is plotted in orange, (D) in yellow and (E) in dark yellow. For both the RS model and the SW model the mass of the first gauge KK mode will be about two times $M_{\rm{KK}}$. Note plotted here are the average values. Typically one can always find tuned points, in parameter space, that satisfy the experimental constraints for all configurations and all KK scales considered. $\Omega=10^{15}$.      }
\label{ fig:epK}
\end{center}
\end{figure}

\begin{figure}[ht!]
    \begin{center}
        \subfigure[]{%
           \label{Fig:NoEpkPoi}
           \includegraphics[width=0.64\textwidth]{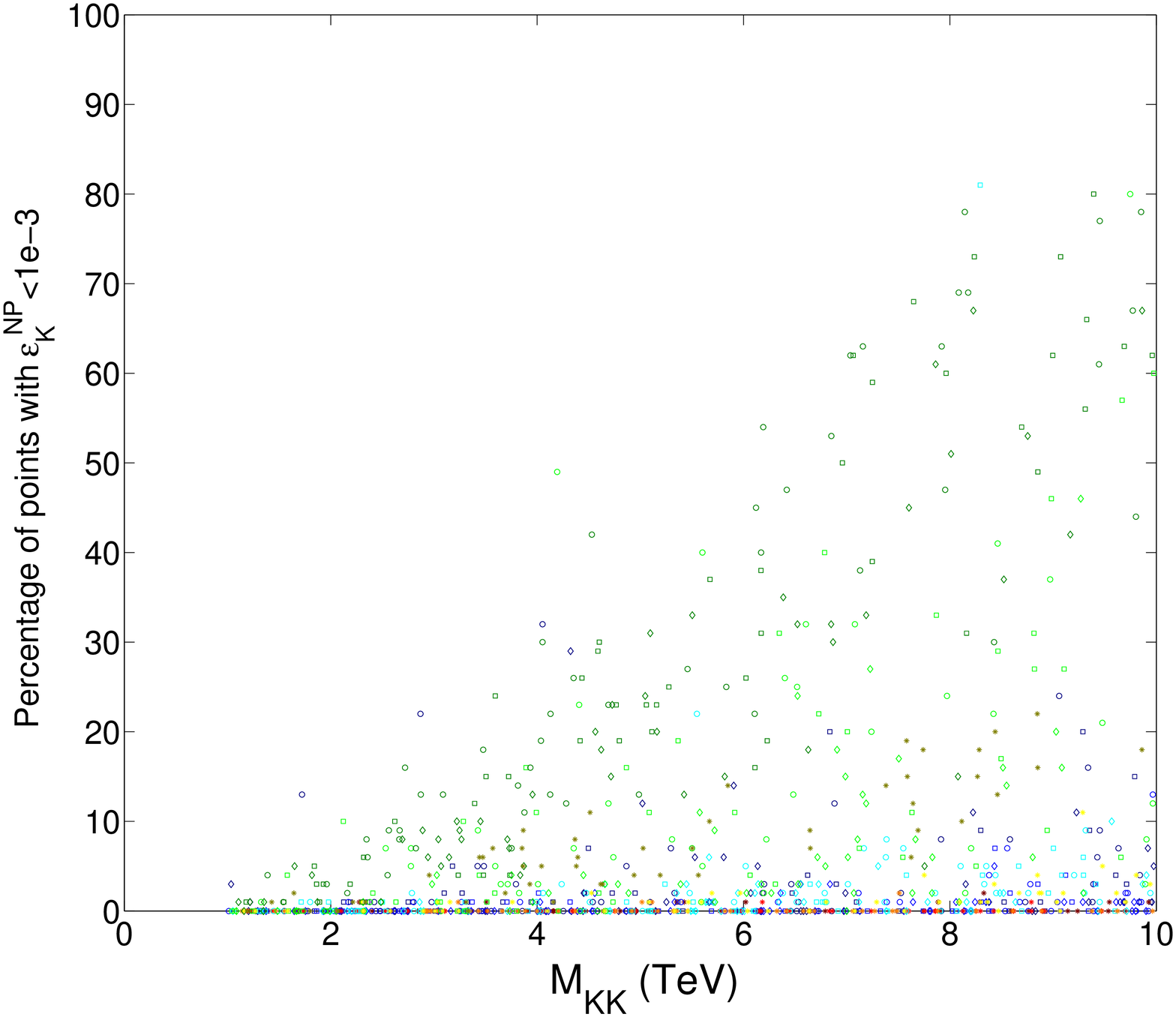}
        }
        \subfigure[]{%
           \label{Fig:NoMkPoi}
           \includegraphics[width=0.64\textwidth]{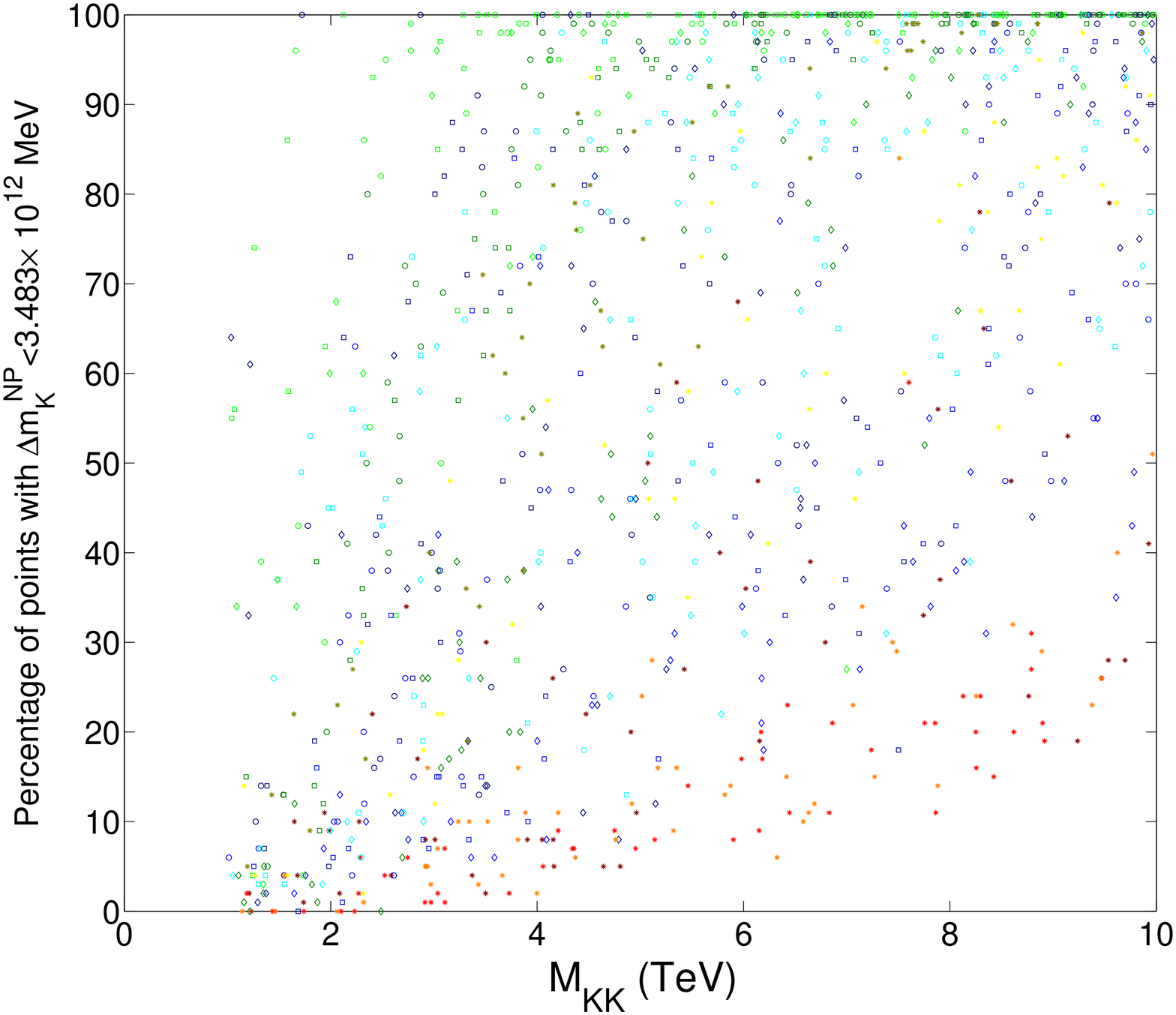}
        }
         \end{center}
    \caption{The percentage of points that are consistent with experimental values. The colours and points correspond to those used in figure \ref{ fig:epK}. Due to the fact that only a 100 points were used, a large number of the configurations and KK scales considered yielded either one or zero points consistent with $\epsilon_K$ (figure \ref{Fig:NoEpkPoi}). Such configurations would require a significant level of tuning in order to be consistent with experimental constraints.} \label{fig:NumPoints}
\end{figure}

\begin{figure}[t!]
\begin{center}
\includegraphics[width=0.8\textwidth]{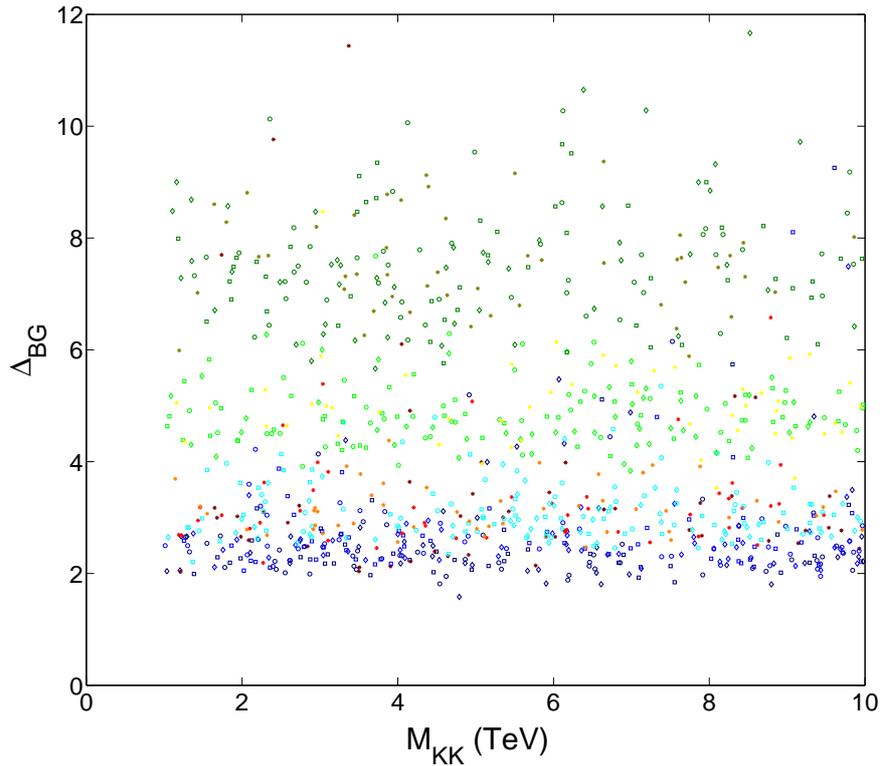}
\caption{The fine tuning parameter giving an indication of the sensitivity of the quark masses, mixing angles and Jarlskog invariant to variations in the Yukawa couplings. The colours and points correspond to those used in figures \ref{ fig:epK}. }
\label{fig:finetune}
\end{center}
\end{figure}

 \begin{itemize}
 \item With the exception of the overall warp factor ($\Omega=10^{15}$) and the KK scale, the only input parameters we fix by hand are the bulk mass parameters $c_L$ and the universal $c_1$ parameter. Here we consider five configurations of $c_L$
  \begin{eqnarray}
(A)\quad c_L=[0.72,\;0.64,\;0.52]\quad\quad\quad(B)\quad c_L=[0.69,\;0.63,\;0.49] \nonumber\\
(C)\quad c_L=[0.66,\;0.60,\;0.42]\quad\quad\quad(D)\quad c_L=[0.63,\;0.57,\;0.34] \nonumber\\
(E)\quad c_L=[0.60,\;0.52,\;0.25]\hspace{3cm}\label{SWquarkConfig}
\end{eqnarray}   
 and three $c_1$ values, $c_1=0.5,\;1,\;1.5$. The five $c_L$ configurations have been chosen such that they give roughly the correct mixing angles. Note that for configurations with $0.74<c_0^{L1}(c_L^1)<0.60$ it becomes increasingly difficult to get a good fit to the quark masses without including quite large bulk mass parameters. 
 \item Next we find the `natural' $c_R^u$ and $c_R^d$ values. By natural we mean the bulk mass parameters that give the correct masses and mixing angles assuming that there is no hierarchy in the Yukawa couplings. For this we generate ten sets of two Yukawa matrices and for each one solve for $c_R^{u/d}$ by fitting to the quark masses. To avoid accidentally using a fairly extreme Yukawa, the $c_R$ values used is the median of these ten values. The Yukawa matrices are generated such that $|\lambda_{ij}|\in[1,\;3]$. As discussed in section \ref{sect:converge} we avoid using large Yukawa couplings. These average $c_R$ values are given in table \ref{Tab:BulkmassParams}.
\item With the nine bulk mass parameters fixed, we proceed to find 100 points in parameter space, that give the correct quark masses, mixing angles and Jarlskog invariant by solving for the Yukawa couplings. The quark masses are run up from $2$ GeV to the mass of the first KK gauge mode ($2M_{\rm{KK}}$) using the $2$ GeV values
 \begin{eqnarray}
m_u=2.5\;\rm{MeV}\quad m_c(3\;\rm{GeV})=0.986\;\rm{GeV}\quad m_t=164\;\rm{GeV}\hspace{0.35cm}\nonumber\\
m_d=4.95\;\rm{MeV}\quad m_s=96.2\;\rm{MeV}\hspace{1.8cm} m_b=4.163\;\rm{GeV}\label{Quarkmasses}
\end{eqnarray}
and the mixing angles are \cite{Nakamura:2010zzi,  Bona:2007vi}
 \begin{eqnarray}
\label{expMixangle}
V_{us}=0.2254\pm0.00065\quad\quad\quad V_{cb}=0.0408\pm0.00045\nonumber\\
 V_{ub}=0.00376\pm0.0002\quad\quad\quad J=2.91^{+0.19}_{-0.11}\;\times 10^{-5}.
\end{eqnarray} 

\item By randomly generating the initial guess of the solver one would anticipate that these points would be spread evenly over the parameter space. However one still needs to check the level of tuning required to obtain masses and mixing angles. Hence here we compute the fine tuning parameter \cite{Barbieri88}  
\begin{equation}
\label{ }
\Delta_{BG}(O_i,p_j)=\left |\frac{p_j}{O_i}\frac{\Delta O_i}{\Delta p_j}\right |,
\end{equation}
where the observables, $O_i$, run over all the masses, mixing angle and Jarlskog invariant and the input parameters, $p_j$, run over the all the Yukawas. In practice we vary the Yukawas over a range of $0.1+0.1i$ (i.e. $\Delta p_j$). The fine tuning parameter is then taken as the maximum value with respect to both input parameters and output parameters. Plotted in figure \ref{fig:finetune} is the mean value taken over the 100 points.  
  
\item Having found these 100 viable points in parameter space we proceed to compute the size of the contributions to $\epsilon_K$ and $\Delta m_K$. The mean value, taken over these 100 points, have been plotted in figure \ref{ fig:epK}. We have also plotted, in figure \ref{fig:NumPoints}, the percentage of the points that are consistent with experimental results. Again in the case of the RS model the KK mass has been scaled by a factor of $\frac{2}{2.45}$ such that the mass of the first gauge KK  mode would be about the same in the two models.

\item This process is then repeated for 50 KK scales, randomly chosen such that $M_{\rm{KK}}\in[1,\;10]$ TeV, for both the RS model and the SW model with $c_1=0.5$, $c_1=1$ and $c_1=1.5$ and also at each configuration in (\ref{SWquarkConfig}).       
 \end{itemize}
 
 As mentioned before, it is hoped that this approach will offer an unbiased spread of points over the region of parameter space of relevance to generating SM fermion masses and mixing angles. Let us turn now to the first question, posed at the beginning of this section, can the SW model generate the correct fermion masses and mixings? As one would expect the answer is yes. For both the RS model and the SW model, not one of the points plotted in figure \ref{fig:finetune} show any sign of significant fine tuning (i.e $\Delta_{BG}>10$). This is of course because of bulk configurations (\ref{SWquarkConfig}) taken as input parameters. Had one chosen more UV or IR localised configurations then clearly more tuning would have been required in order to get the correct masses and mixing angles.   
 
This then brings us to the second question, related to the the suppression of FCNC's. By grouping points in parameter space according to their configurations of bulk mass parameters (\ref{SWquarkConfig}), one is essentially comparing points which require equivalent levels of tuning in order to obtain the correct masses and mixing angles (see figure \ref{fig:finetune}). Even with this comparison it is still difficult to meaningfully quantify the extent to which FCNC's are suppressed. None the less one can see that, for all points considered, FCNC's are considerably more suppressed in the SW model than in the RS model with an equivalent level of tuning. For example, one can see from figure \ref{Fig:NoEpkPoi} that, in the SW model with configuration (E), in order to have about 20\% of the points consistent with $\epsilon_K$ one would require $M_{\rm{KK}}\gtrsim 4$ TeV (corresponding to a KK gluon mass of $\sim 8$ TeV). Where as in the RS model one would require $M_{\rm{KK}}\gtrsim 10$ TeV (corresponding to a KK gluon mass of $\sim 25$ TeV). Alternatively one can look at the total number of points that satisfy the $\epsilon_K$ constraint regardless of KK scale. In the SW for configuration (E) with $c_1=1.5,\;1,\;0.5$ this is about $31\% ,\; 31\% ,\; 19\%$ respectively. While for configuration (A) this is about  $4.2\% ,\; 2.5\% ,\; 2.6\%$. This can be compared to the RS model which is about $8\%$ for configuration (E) and $0.22\%$ for configuration (A).           

There are a number of factors contributing to this increased suppression of FCNC's. The extent to which FCNC's are suppressed is largely determined, at tree level, by how universal the gauge-fermion couplings, that appear in (\ref{fourFermInt}), are. As can be seen in figure \ref{fig:FermCoupl}, as one localises the fermions further and further towards the UV, the gauge fermion coupling becomes increasingly universal. While this is true for both the SW model and the RS model one can also see, from figure \ref{fig:UVFermCoupl}, that in the SW model, with $c_1\gtrsim 1$ that the gauge fermion couplings are slightly more universal than the RS model. In addition to this in the SW model, due to the presence of the bulk Higgs, the fermions will typically sit slightly further towards the UV than in the RS model (see table \ref{Tab:BulkmassParams}). This effect should be combined with the reduced $c_0$ dependence in the range of possible fermion masses (the gradient in figure \ref{fermmass}). This results in avoiding extreme bulk mass parameters in configurations in which either the left-handed or right-handed fermions are localised quite far towards the UV. For example, see $c_R^{u3}$ for configurations (A) and (B) in table \ref{Tab:BulkmassParams} for an extreme example. When all these effects are combined one finds that, for all points in parameter space considered, FCNC's are more suppressed in the SW model than in the RS model. 

One of the underlying assumptions in this analysis is that of a universal $c_1$ value. However, upon examining the results one can see that, despite most of the bulk mass parameters of relevance to kaon physics having $c_0^R\lesssim-0.6$, there is still a very small $c_1$ dependence. In particular smaller $c_1$ values tend to give slightly more UV localised fermions and slightly less universal gauge fermion couplings. These effects are quite small and hence negligible when compared to varying the bulk mass parameters, $c_0$. So here we would argue that these results would not change significantly if one was to relax this assumption. None the less, since the exact origin of the $c_1$ term has not been clearly defined it is not clear if this number can be quite large or small. Here we shall leave investigation of this to future work.     

It should also be stressed that this cannot be considered a complete study for a number of reasons. Firstly we only consider tree level gauge mediated FCNC's. One would also anticipate additional contributions to FCNC's arising from, for example, the dilaton or the Higgs \cite{Casagrande:2008hr, Azatov:2009na, Agashe:2009di, Buras:2010mh}. The inclusion of such additional contributions would inevitably enlarge the parameter space and hence make a fair comparison with the RS model more difficult. However one would anticipate that such fields would be IR localised and hence such FCNC's should also be suppressed. Secondly we also only focus on kaon physics. One can see from table \ref{Tab:BulkmassParams} that a study involving top physics or B physics would involve fermions sitting further towards the IR where the assumption of universal $c_1$ values is arguably less valid. So here we will leave a more comprehensive study to future work.   

It is important to realise that the central physics involved in this result is not necessarily related to the soft wall but rather the change in the relationship between fermion masses and their positions. In particular the reduction of the gradient in figure \ref{fermmass}. One would anticipate such an effect showing up, to some extent, in any bulk Higgs scenario. For example, a similar reduction, in the constraints from $\epsilon_K$, was found in the RS model with a bulk Higgs in \cite{Agashe:2008uz}. However soft wall models offer a framework in which the Higgs can propagate in the bulk and the gauge hierarchy problem can still potentially be resolved.  

%%%%%%%%%%%%%%%%%%%%%%%%%%%%%%%%%%%%%%%%%%%%%%%%%%%%
\section{Discusion and Conclusions}\label{sect:conclusion}

The primary motivation of this paper was to demonstrate that the description of flavour that exists in the RS model can be transferred across to the SW model. Here we believe that we have done this but there is one major difference. Whereas in the RS model, with the Higgs on the brane, the fermion masses are determined by the end points of the fermion profiles, in the SW model they are determined by the overlap integral of the Higgs VEV and the fermion profiles. This gives rise to two significant differences between the SW model and the RS model. Firstly, although the range of possible masses still falls exponentially, in the SW model the exponent is no longer constant. Secondly, with both left-handed and right-handed fermions sitting towards the UV one obtains a minimum possible fermion mass. 

Both the gradient in the fermion masses and the minimum fermion mass are very sensitive to both the warp factor and the form of the Higgs VEV. One would also suspect that it would be sensitive to the form of the dilaton background value as well. Although here we have fixed $\nu=2$ in order to arrive at a Regge scaling in the KK masses. It should also be noted that one cannot obtain the hierarchy of fermion masses if the warp factor is too small or the Higgs VEV is too flat. However, with a warp factor of $\Omega=10^{15}$ the case of a quadratic Higgs VEV is particularly interesting since it results in a minimum fermion mass being approximately fifteen orders of magnitude lower than the EW scale. This appears to be in rough agreement with the observed range of neutrino masses and hence the majority of this paper has focused on this case. Further still it was found in \cite{Cabrer:2011fb} that an approximately quadratic Higgs VEV gave the lowest EW constraints. 
 
Having restricted the study to that of a quadratic Higgs VEV we proceeded to investigate the implications for flavour physics. Studies of flavour in the SW model suffer from a number of additional complications to that of the RS model. Firstly, one must necessarily ensure that the bulk fermions gain a $z$-dependent mass term in order to obtain a discrete KK spectrum. The most natural source of such a mass term is the Yukawa couplings to the Higgs. However here, following \cite{MertAybat:2009mk},  we use a generic bulk mass (\ref{bulkmass}) term which could or could not be related to the Yukawa couplings. The advantage to this bulk mass term is that it allows for analytical expressions to be obtained for the fermion profiles and hence simplifies the study. The disadvantage is that one does not know the precise relationship between the $c_1$ parameters and the Yukawa couplings. Fortunately it is found that when the fermions are localised towards the UV and one assumes $c_1\sim\mathcal{O}(1)$ then the results are relatively independent of $c_1$. Hence for the physics considered here, notably kaon physics and lepton decays, one can assume a universal $c_1$ value and still arrive at a reasonably reliable result.   

The situation receives an additional complication when one tries to estimate at what scale one loses perturbative control of the calculation. With the KK masses scaling as $m_n^2\sim n$, one would naively expect a tree level sum over the KK tower to be logarithmically divergent. I.e. tree level processes are dominated by the higher KK modes and hence one must impose a cut off in the KK number. By computing the five dimensional gauge propagator, it is demonstrated that this divergence does not occur provided the fermions gain a sufficiently $z$-dependent mass term, i.e. $c_1\gtrsim 0.5$. If one assumes that the Yukawa couplings are the source of the $c_1$ term and that $c_1\sim \lambda h_0$ then this would suggest that one needs large Yukawa couplings to avoid such a tree level divergence. However this would result in a loss of perturbative control, at next to leading order,  at a scale much lower than the KK scale. Alternatively one could look for an another source of the $z$-dependent mass term such as couplings to the dilaton or
a Goldberger-Wise scalar \cite{Medina:2010mu}. 

None the less, assuming $c_1\sim\mathcal{O}(1)$ and a quadratic Higgs VEV, we are then able to investigate the phenomenological implications of the change in the range of possible fermion masses. Due to the difficulties in quantifying the extent to which FCNC's are suppressed here we choose to compare the SW model to the well studied RS model. In other words, we are using the RS model as a point of reference. Firstly, a relatively simple study of rare lepton decays is made primarily to check the assumption of universal $c_1$ values as well as allowing for comparison with some earlier work \cite{Atkins:2010cc}.  With $c_1\approx 1$, good agreement is found with models in which the backreaction of the Yukawa couplings, on the fermion profiles, is included. Next we proceed to examine $K^0-\bar{K}^0$ mixing for both the RS model and the SW model, comparing points in parameter space that require equivalent levels of tuning in order obtain the correct masses, mixing angles and Jarlskog invariant. It is found that, in the SW model, the fermions typically sit slightly further towards the UV than in the RS model, with a brane localised Higgs. Also the gauge fermion couplings are slightly more universal. This results in FCNC's being more suppressed in SW model than in the RS model for nearly all points in parameter space considered. Hence here we would conclude that, despite its computational difficulties and uncertainties, the SW model arguably offers a more appealing description of flavour than the RS model. This result, coupled with the result of \cite{Cabrer:2011fb, Cabrer:2010si}, suggests that there is a strong case to be made for considering models with a bulk Higgs. 
   
\section*{Acknowledgements}
We would like to thank Michael Atkins for collaboration in the early stages of this work.
P.R.A.~was supported by the Science and Technology Facilities Council. 
S.J.H.~was supported by the Science and Technology Facilities Council [grant number ST/G000573/1]. S.J.~was supported by the Science and Technology Facilities Council [grant number ST/H004661/1] and acknowledges support from the NExT institute and SEPnet.   
   
\bibliographystyle{JHEP}

\bibliography{SWbibliography}   

\end{document}